\documentclass[%
 reprint,
superscriptaddress,
nofootinbib,
nobibnotes,
 amsmath,amssymb,
 aps,floatfix,
]{revtex4-2}

\usepackage{amsthm}
\usepackage{graphicx,xcolor}
\usepackage{dcolumn}
\usepackage{bm}
\usepackage[breaklinks=true,colorlinks,citecolor=blue,linkcolor=blue,urlcolor=blue]{hyperref}
\usepackage[mathlines]{lineno}
\usepackage[caption=false, position=bottom]{subfig}
\usepackage{floatrow}
\usepackage{mathtools}

\usepackage[utf8]{inputenc}
\usepackage[english]{babel}

\newtheorem{theorem}{Theorem}

\newcommand{\bp}{\boldsymbol{\epsilon}}
\newcommand{\si}{\boldsymbol{\Sigma}}

\newcommand{\br}{\bar{\rho}}
\begin{document}


\preprint{APS/123-QED}

\title{Correlation Thresholds For Effective Composite Pulse  Quantum Error Mitigation}

\author{Ido Kaplan}%
\affiliation{
School of Electrical Engineering, the Iby and Aladar Fleischman Faculty of Engineering, Tel-Aviv University, Tel-Aviv 69978, Israel.
 }
 


\author{Haim Suchowski}%
\affiliation{
 Raymond and Beverly Sackler School of Physics and Astronomy, Tel-Aviv University, Tel-Aviv 69978, Israel.
 }

\author{Yaron Oz}%
\affiliation{
 Raymond and Beverly Sackler School of Physics and Astronomy, Tel-Aviv University, Tel-Aviv 69978, Israel.
}

\date{\today}

\begin{abstract}

Composite pulse segmentation has emerged as a promising error mitigation technique for a wide range of physical systems. 
In recent years, composite scheme were applied  as mitigation strategies for quantum information processing and quantum computing. However, most of these strategies assume full error correlation between segments, which can result in gates with worse fidelity performance compared to non-composite gates. In our research, we investigate how error correlations impact the fidelity of quantum gates within the composite segmentation framework. In our study, we prove the existence of a critical correlation threshold, above which the composite pulse method significantly enhances both the mean value and variance of the fidelity.
To gain deeper insights, we analyze various properties of the threshold in the realm of integrated photonics, including the effects of geometrical variations and the limit where the number of segments approaches infinity. We numerically explore diverse scenarios, showcasing different aspects of the critical threshold within the photonic quantum gates framework. These findings contribute open new pathways of error mitigation strategies and their implications in quantum information processing.

\end{abstract}

\maketitle
\textit{Introduction.}\textbf{—}Composite pulse (CP) segmentation, a technique where series of pulses interact sequentially in order to minimize systematic errors, has been employed as an error mitigation scheme
in diverse classical and quantum systems \cite{Levitt1979,Shaka1985,Shaka1987,Levitt1986,Timoney2008}. While a standard 
assumption in the composite-segmentation framework is that there is full correlation
between the errors of the different segments \cite{Torosov2022}, in reality, full correlation between these errors is not generally the case, and the errors can be partially correlated \cite{partialCorrelated} or even completely uncorrelated \cite{nation2021scalable, Preskill_1998}.
It is intuitively clear that a lack of correlation between the errors will make the segmented pulse
less efficient than a single pulse (or segment) because of the accumulation of the independent errors.
One should thus expect that there is a correlation threshold above which CP segmentation is effective, which is reminiscent of the threshold of error correction codes.

The aim of this paper is to analyse this correlation threshold and 
the effect of error correlations on the performance 
of the CP segmentation applied to quantum gates.
We will prove that there is a critical correlation threshold above which the mitigation method
increases the fidelity. We will study different properties of the
threshold including  the mean and variance of the fidelity, the threshold's geometrical shape and its behaviour in the limit of a large number
of segments.
We provide diverse numerical simulations for photonic quantum gates \cite{kyoseva2019detuning,YaronIdo}
with systematic errors that exhibit different aspects of the critical threshold.

The paper is organized as follows. In section 'Correlated Error Mitigation - Theoretical Claims' we present the threshold theorems and outlined proofs.
In the following sections afterwards we  provide examples to the theorems using numerical simulations of photonic quantum gates.
The final section is devoted to a discussion and outlook. 
We detail the proofs of theorems and the numerical simulations in the supplemental material.

\textit{Correlated Error Mitigation - Theoretical Claims.}\textbf{—}
Consider an ideal unitary gate, denoted as $U_{ideal}$, and its actual physical realization, $U_{phy}(\boldsymbol{\epsilon})$. The errors are modelled as a set of variables $\bp= \{\epsilon^a\}_{a=1}^m$ which we estimate as jointly distributed random variables.
We will study the gate fidelity $F$, which is a function  of $U_{ideal}$ and $U(\bp)$: 
\begin{equation}
F(U_{ideal},U(\bp))= \frac{1}{2}\cdot |Tr(U_{ideal}^{\dagger}U(\bp)| \ ,
\label{GF}
\end{equation}
that takes values in the interval $[0,1]$.

The composite pulses is a framework for error mitigation \cite{torosov2022experimental} where $U(\bp)$
is realized as a product of $n$ physical unitary gates:
\begin{equation}
U_{cp}(\vec{\bp})= \prod_{i=1}^n U_i(\bp_i)  \ ,
\label{CP}
\end{equation}
where $\vec{\bp} = (\bp_1,...,\bp_n)$, $\bp_i = \{\epsilon^a_i\}_{a=1}^m$, and $U_i$ are constructed such that the fidelity (\ref{GF}) is higher than the physical fidelity. 
For simplicity and for a practical application to systematic errors in
the fabrication process, we will assume that the errors are are jointly distributed Gaussian
random variables with zero mean. We will further assume that the dimensionless variance of the error is small
$\sigma^2 \ll 1$, which is typically the case in photonic quantum gates.
The generalization of our analysis to other distributions and larger variance is straightforward. 

We expand the physical fidelity and CP fidelity in power series:
\begin{eqnarray}
F_{phy}(\bp) = 1 + b_{ab}\epsilon^a \epsilon^b + c_{abcd}\epsilon^a \epsilon^b\epsilon^c \epsilon^d + O(\epsilon^6) 
\nonumber\\
F_{cp}(\vec{\bp}) = 1 + b_{ab}^{ij}\epsilon^a_i \epsilon^b_j + c_{abcd}^{ijkl}\epsilon^a_i 
\epsilon^b_j\epsilon^c_k \epsilon^d_l + O(\epsilon^6)  \ , 
\label{FR}
\end{eqnarray}
and we use Einstein's summation convention.
In the absence of errors both fidelities (\ref{FR}) have the same value, one. 
We will consider the mean and variance of the fidelities.


In the standard CP scheme one assumes that for every $i=1,...,n,a=1,...,m$
the errors $\{\epsilon_i^a\}$ 
are fully correlated.
This is generally not the case in real life applications, where the degree of correlation can
vary. It is therefore natural to ask what are the implications of partial correlation on the performance of the CP mitigation scheme. It is qualitatively clear that 
weakly correlated errors can make the CP ineffective. In particular, in the extreme case of uncorrelated errors the CP scheme will decrease the fidelity compared to a single pulse due
to the accumulation of the independent errors. In the following theorems we quantify the performance of CP as a function of the degree of correlation between the errors. We will consider the mean and the variance of the fidelities.
We will use the notations $\delta_{ij}, \delta^{ab}$ for the Kronecker delta-function and $H_{ij}=1, \forall i,j =1,...,n$,
$H^{ab}=1, \forall a,b=1,...,m$. It will be convenient to use the matrices
$D_{ij} = H_{ij}-\delta_{ij}$ and $D^{ab} = H^{ab}-\delta^{ab}$.

\begin{theorem}
Given a composite pulse, described in Eq. (\ref{CP}), where the errors of each segment are fully correlated but 
the errors of different segments are
partially correlated, there is a critical correlation 
above which the mean of the composite pulse fidelity is higher than the mean of the physical 
fidelity.

\end{theorem}

\begin{proof}
We have that $\{\epsilon_i^a\}_{a=1}^m$ are fully correlated for
$i=1,..,n$, but  $(\bp_i, \bp_i),i\neq j$ are
partially correlated.
The covariance matrix reads:
\begin{equation}
\Sigma^{ab}_{ij} = \mathbf{E}[\epsilon_i^a \epsilon_j^b] = \sigma^2H^{ab}(\delta_{ij} + \rho D_{ij}) \ ,  
  \label{segerrors_ij}
\end{equation}
where $\rho\in[0,1]$ and $\sigma^2 \ll 1$.
The expectation value of the physical and composite fidelities (\ref{FR}) can be written as:
\begin{eqnarray}
\mathbf{E}[F_{phy}]&\equiv& \bar{F}_{phy}  = 1 + b_{phy} \sigma^2 +  
O(\sigma^4)\nonumber\\
\mathbf{E}[F_{cp}]&\equiv& \bar{F}_{cp} 
= 1 + b_{cp} \sigma^2 + c_{cp} \rho\sigma^2 + O(\sigma^4) \ ,
\label{meanfp}
\end{eqnarray}
where $b_{phy} = H^{ab}b_{ab}$,
$b_{cp} = H^{ab}\delta_{ij}b_{ab}^{ij}, c_{cp} = H^{ab}D_{ij}b_{ab}^{ij}$.
Equating $\bar{F}_{phy}= \bar{F}_{cp}$ at the critical correlation $\rho_c$ we find:
\begin{equation}
    \rho_c = \frac{b_{phy}-b_{cp}(n)}{c_{cp}(n)} \ ,
    \label{cor_between_segments}
\end{equation}
where we wrote explicitly the dependence on the number of pulses $n$.
$b_{phy} < 0$ and $b_{cp} < 0$ since the fidelity decreases when increasing the error variance, while
$c_{cp} > 0$ since higher correlation $\rho$ increases it.
Note also that $\rho_c$ is independent of the standard deviation in  Eq. (\ref{cor_between_segments}).
$\bar{F}_{cp}$ is a monotonically increasing function of $\rho$ and is greater
than $\bar{F}_{phy}$ for $\rho > \rho_c$.
When $\rho_c > 1$ it means that there is no region where CP improves
the fidelity of the gate.

\end{proof}

\noindent
{\it Remarks}\\
(i) $b_{cp}$ and $c_{cp}$ in (\ref{meanfp}) get contributions from $n$ 
and $n(n-1)$ terms, respectively. Thus, 
the scaling of the critical correlation (\ref{cor_between_segments}) at large $n$ is:
\begin{equation}
\rho_c  \sim \frac{1}{n} \ , 
\label{scaling}
\end{equation}
which implies that the larger the number of segments, the lower the threshold correlation.


\noindent
(ii) Given a CP solution for $\rho=1$ we can ask how effective this segmented configuration is when we decrease $\rho$.
We have:
\begin{equation}
    \frac{\bar{F}_{cp}(\rho) - \bar{F}_{phy}(\rho)}{\bar{F}_{cp}(1) - \bar{F}_{phy}(1)} = \frac{b_{cp} + c_{cp} \rho - b_{phy}}{b_{cp} + c_{cp} - b_{phy} } \leq 1 \ ,
\end{equation}
where $b_{cp}$ and $c_{cp}$ are the coefficients of the $\rho=1$ solution, and $\rho_c$ is given by (\ref{cor_between_segments})
evaluated on this solution.
We will see that given $\rho < 1$ one can find other segmented solutions with a lower $\rho_c$, which however
will have a lower fidelity than that of the $\rho=1$ solution when applied to the fully correlated case.
We will define $\rho_c$ to be the minimal value on the space of solutions.

\begin{theorem}
If the errors of different segments are fully correlated
but the errors of each segment are partially correlated, 
there is a critical correlation 
above 
which the mean of the composite pulse fidelity is higher than the mean of the physical 
fidelity.

\end{theorem}

\begin{proof}
We have that $(\bp_i,\bp_j), i\neq j$ are fully correlated
but $\{\epsilon_i^a\}_{a=1}^m$ for any fixed $i$ are partially correlated.
The covariance matrix in this case reads:
\begin{equation}
\si^{ab}_{ij}= \sigma^2H_{ij}(\delta^{ab} + \rho D^{ab}) \ ,
\label{cov2}
\end{equation}
where $\rho\in[0,1]$ and $\sigma^2 \ll 1$. 
 The critical correlation $\rho_c$ is:
\begin{equation}
    \rho_c = -\frac{b_{phy}-b_{cp}}{c_{phy}- c_{cp}} \ ,
    \label{cor_between_waveguides}
\end{equation}
where unlike (\ref{cor_between_segments}), here $b_{cp}$  and $c_{cp}$ are independent of the number of pulses $n$.
$\bar{F}_{cp}$ is a monotonically increasing function of $\rho$ and is greater
than $\bar{F}_{phy}$ for $\rho > \rho_c$.

\end{proof}
The above two theorems can be combined to the case where both $\{\epsilon_i^a\}_{a=1}^m$ and $(\bp_i,\bp_j), i\neq j$ are correlated. In such a case there is a critical curve of correlations that
separates the regime $\bar{F}_{cp} > \bar{F}_{phy}$ from $\bar{F}_{cp} <  \bar{F}_{phy}$.

\begin{theorem}
Given a composite pulse (\ref{CP}), there is a critical correlation curve
separating the effective and non-effective error mitigation regimes.

\end{theorem}

\begin{proof}
The covariance matrix takes the form: 
\begin{eqnarray}
 \si^{ab}_{ij} = \sigma^2(\delta_{ij}\delta^{ab} + \bar{\rho} \delta_{ij}D^{ab} +
 \rho \delta^{ab}D_{ij} +\rho\bar{\rho}D^{ab}D_{ij})  \ ,
 \label{covgen}
\end{eqnarray}
where $\rho,\bar{\rho}\in[0,1]$ and $\sigma^2 \ll 1$. 
We find that the critical curve ${\cal C}$ is given by:
\begin{eqnarray}
&{\cal C}: E\rho \bar{\rho} + D \rho + C \bar{\rho}  + B  = 0 \ ,
\label{corline1}
\end{eqnarray}
and region where the composite pulse is effective is the area  bounded 
by the curve ${\cal C}$ and the lines $\rho=1$ and $\bar{\rho}=1$.
The details of the proof are in the supplemntal material. Note, that when the standard deviation increases we may need to expand the fidelities to higher powers and the curve ${\cal C}$ becomes a higher degree polynomial.
\end{proof}

In the following we inquire whether there is a relationship between the
region where the mean CP fidelity is higher than the physical one, and the corresponding region 
for the standard deviations of the fidelity.
\begin{theorem}
The critical correlation for the standard deviation of the fidelity is higher than that of the 
mean fidelity.
\end{theorem}

\begin{figure}[!h]
     \centering
     \includegraphics[width=6.5cm]{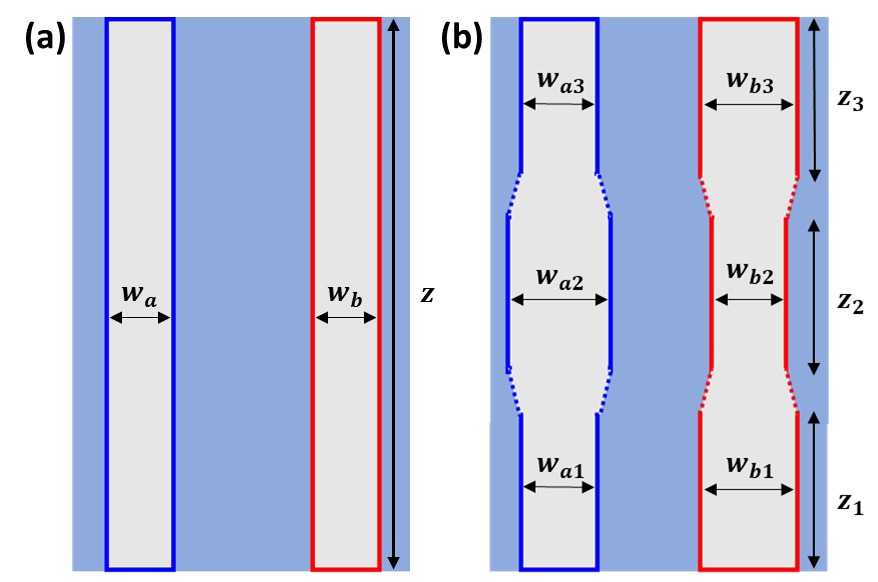}
     \caption{Illustration of two types of DCs. In figure (a) we see a uniform DC, which is charaterized by three geometric parameters - the width of the waveguides $w_a,w_b$, and the DC's length $z$. The geometric parameters might have systematic errors in their dimensions. In figure (b) we see a CP-based directional coupler. It is segmented into three parts, where each segment has its own length, and each waveguide in each segment has its own width.
     }
     \label{fig: Couplers illustration}
\end{figure}

\begin{proof}
Consider the covariance matrix (\ref{segerrors_ij}) and denote the mean and standard deviations of the fidelities
(\ref{FR}) by $\mu \equiv \mathbf{E}[F], \sigma \equiv \sqrt{\mathbf{E}[F^2] - \mathbf{E}[F]^2}$.
We prove in the supplemental material that $\mu_{cp} \geq \mu_{phy}$ implies 
\begin{equation}
\sigma_{cp} \leq \sqrt{2}(1-\mu_{cp}) \leq \sqrt{2}(1-\mu_{phy}) \leq  \sigma_{phy} \ .   
\end{equation}

\end{proof}
\textit{Correlated Error Mitigation - Numerical Simulations.}\textbf{—}In this section we will perform explicit simulations of the CP error mitigation
for photonic quantum gates and demonstrate numerically the threshold theorems of the previous section.
Photonics based quantum information processes (QIPs) utilize photons as efficient
low-noise carriers \cite{O_Brien_2009} of quantum information. 
The qubit in dual rail QIPs is constructed by using a single photon and two waveguides; if the photon is 
in the first waveguide, the qubit is in the state $|0\rangle$ and if it is in the second waveguide, the qubit is in the
state $|1\rangle$. In the supplemental material \ref{appendix: DC structure}, we provide more details regarding the directional coupler (DC) structure.
A segmented DC is plotted in figure \ref{fig: Couplers illustration}, where
the parameters quantify the geometric properties of the DC.

In all the simulations presented in this paper, we estimated the standard deviation to be $3 \sigma = 20[nm]$, which is about $5$ percent of the waveguides' width - common fabrication variation value \cite{Ding_13}. Hence the dimensionless standard devitations are $\sigma \sim 0.166 \ll 1$.
The optimization process itself is explained in further detail in Refs. \cite{YaronIdo}.   

\begin{figure}[!h]
     \includegraphics[width=8cm]{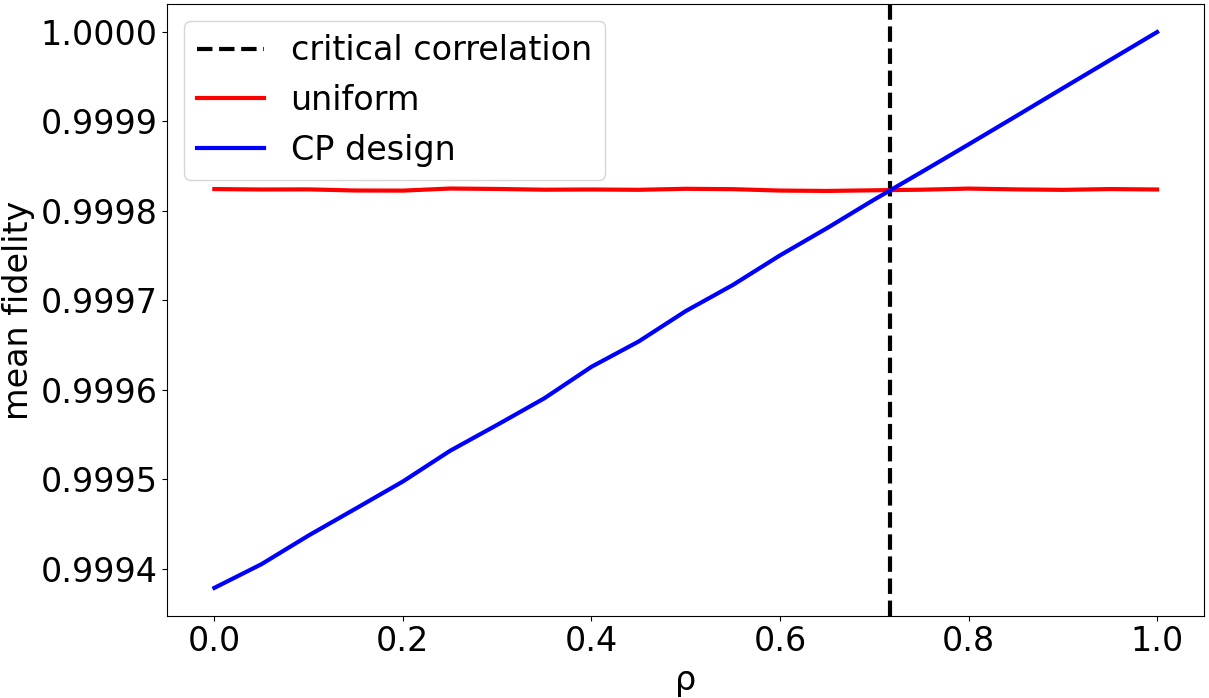}
     \caption{Mean fidelity as a function of $\rho$ for a fixed $\sigma \sim 0.166 \ll 1$. This plot is the $\rho=1$ solution applied
     to lower correlation. The vertical dashed line is the threshold correlation (\ref{cor_between_segments}), and CP is effective above it.}
     \label{fig: fidelity vs rho}
\end{figure}

\newpage
\textit{Partially Correlated Errors Between Different Segments  (Theorem 1)}\textbf{—}Consider a 3-segments pulse with covariance matrix as seen in Eq.
(\ref{segerrors_ij}), 
with the parameter values used for the simulations are given in table \ref{table: rho_crit vs sigma}.
In Figure \ref{fig: fidelity vs rho} we plot the mean fidelity as a function of the correlation parameter $\rho$ for
a fixed standard deviation $\sigma$.
The mean fidelity grows linearly in $\rho$, 
and dashed line in the plot denotes $\rho_{c}$ (\ref{cor_between_segments}).
In the supplemental material \ref{appendix: simulations theorem 1} we present the graph from figure \ref{fig: fidelity vs rho} for various values of $\sigma$, which shows that $\rho_{crit}$ is independent of $\sigma$.

\begin{figure}[!h]
     \centering
     \includegraphics[width=8cm]{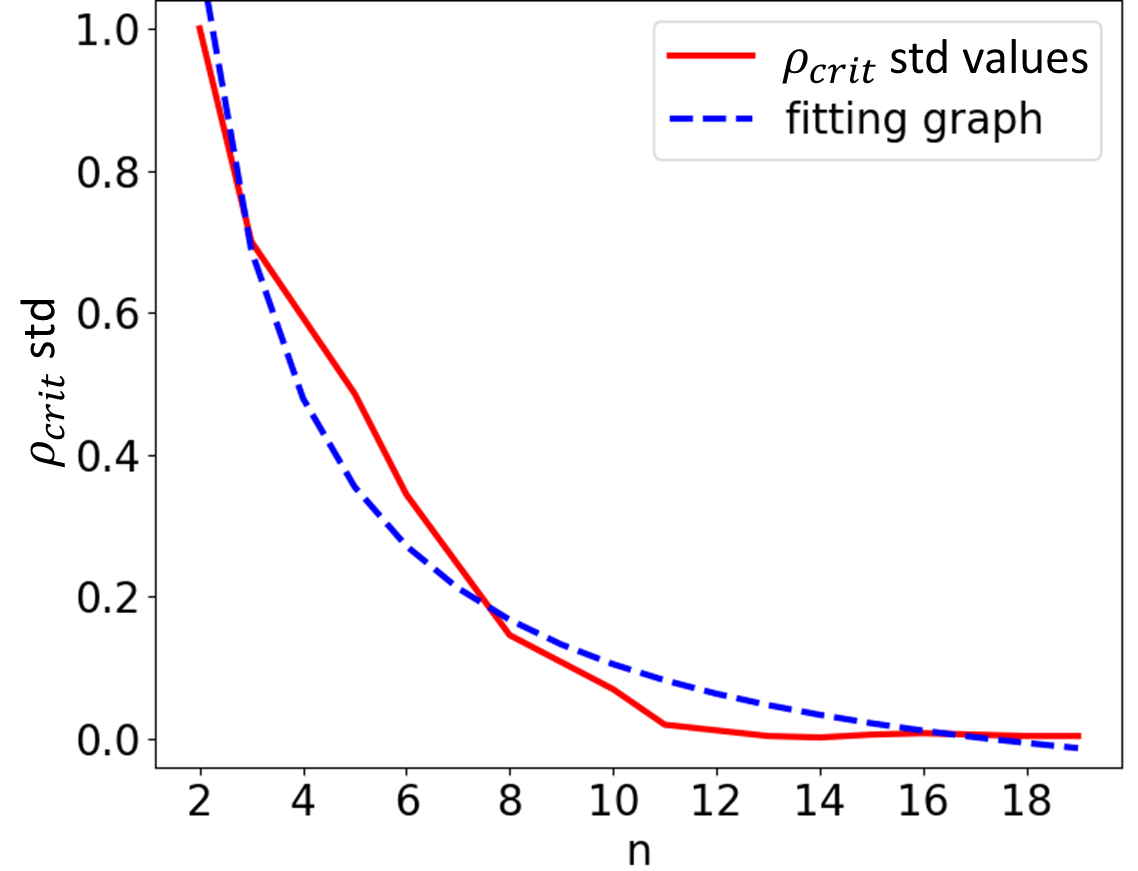}
     \caption{The critical correlation versus the number of segments $n$ exhibiting the $\frac{1}{n}$ scaling (\ref{scaling}).}
     \label{fig: rho crit vs n}
\end{figure}

In Figure \ref{fig: rho crit vs n} we plot the dependence of $\rho_{crit}$ on the number of segments $n$.
We see that the plot is consistent with prediction (\ref{scaling}).
Thus, more segments increase the gate robustness at lower correlation.
The solutions used in order to generate the numerical values in figure \ref{fig: rho crit vs n} are presented in appendix \ref{sets of solutions for rho_crit vs n}.

\begin{figure}[!h]
     \centering
     \includegraphics[width=7.7cm]{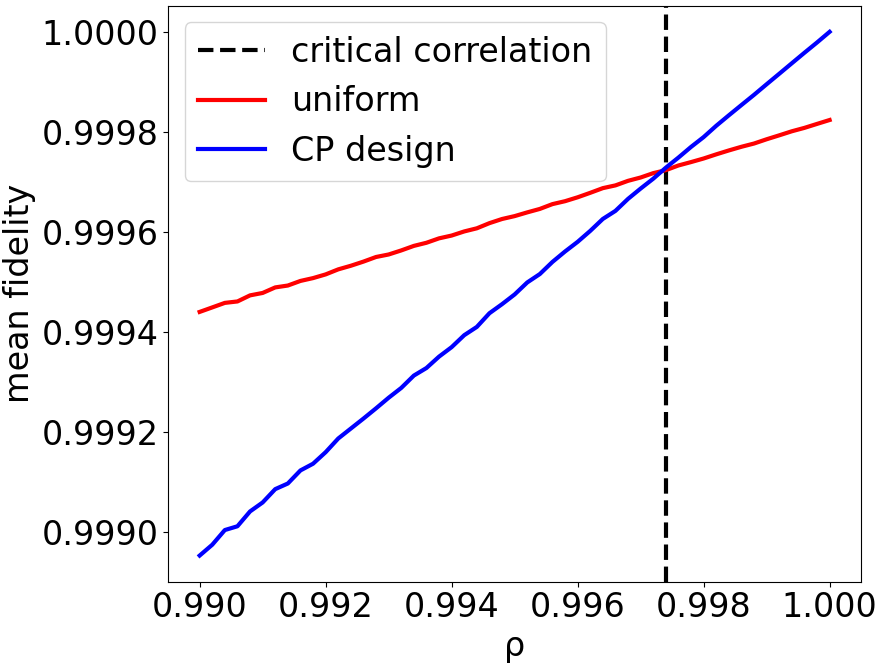}
     \caption{Mean fidelity as a function of $\rho$ for a fixed $\sigma$. This plot is the $\rho=1$ solution applied
     to lower correlation. The vertical line is the threshold correlation (\ref{cor_between_waveguides})
     and is higher than (\ref{cor_between_segments}).}
     \label{fig: rho between waveguides vs fidelity}
\end{figure}

\newpage
\textit{Partially Correlated Errors of Individual Segments  (Theorem 2)}\textbf{—}In the following simulations we will consider the 3-segments pulse of the previous section with covariance matrix
(\ref{cov2}). In Figure \ref{fig: rho between waveguides vs fidelity} we plot the mean fidelity as a function of $\rho$ 
for a fixed standard deviation $\sigma$ and we see the linear growth (\ref{Lin}).
Note that in Figure \ref{fig: rho between waveguides vs fidelity} the critical correlation parameter $\rho_c$
is much larger than in Figure \ref{fig: fidelity vs rho}.
Thus, while the segmented solutions are robust even with partial correlations between different segments, this
is not the case when the two waveguides are partially correlated within the same
segment. The effect of correlation between $\delta w_1$ and $\delta w_2$ is stronger than between segments and
even a small reduction in the correlation in the latter case has a significant impact on the fidelity.
In supplemental material \ref{appendix: simulations theorem 2} we present the graph from figure \ref{fig: rho between waveguides vs fidelity} for various values of $\sigma$, which shows that $\rho_{crit}$ is independent of $\sigma$. Furthermore, we present in  supplemental material \ref{appendix: simulations theorem 2} how the critical correlation line is independent of the number of segments as well.

\begin{figure}[!h]
     \centering
     \includegraphics[width=8cm]{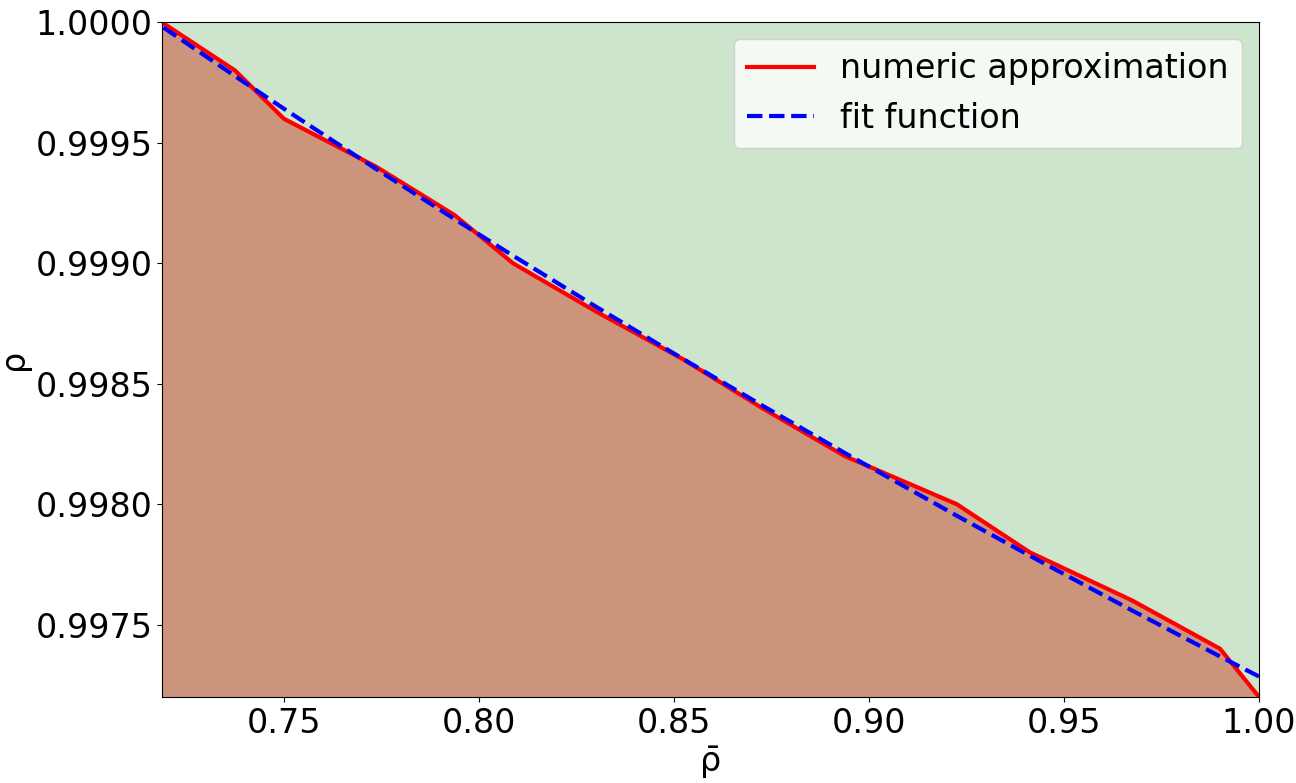}
     \caption{The critical curve between (\ref{corline}) separating the effective (green) from the non-effective (red) mitigation regimes.
     The vertical axis denotes the correlations between the different segments. The horizontal axis denotes the correlations between the two wave-guides of a given segment. The fit function for the curve is detailed in equation \ref{corline} and the constant values used for the equation are 
$E = 0.945, 
B = -1.347, 
C = -0.923, 
D = 1.331
$.}
     \label{fig: critical line}
\end{figure}

\newpage
\textit{Correlation Curve (Theorem 3)}\textbf{—}With correlations parameters $(\rho,\br)$ between different segments and between the 
waveguides in the same segment we found a critical line separating the effective and non-effective error mitigation regimes (\ref{corline}).
Using the parameters from table 1, and a fixed standard deviation $\sigma$, we analyzed numerically the critical line in the $(\rho,\br)$ plane, which is indeed a polynomial function
in Figure \ref{fig: critical line}.
The numerical approximation, seen in Figure \ref{fig: critical line}, of the critical curve is noisy, however, it's very similar to the polynomial estimation function, with maximal difference of only $1\% $, and mean difference smaller than $0.5\% $.

\begin{figure}[!h]
     \centering
     \includegraphics[width=8cm]{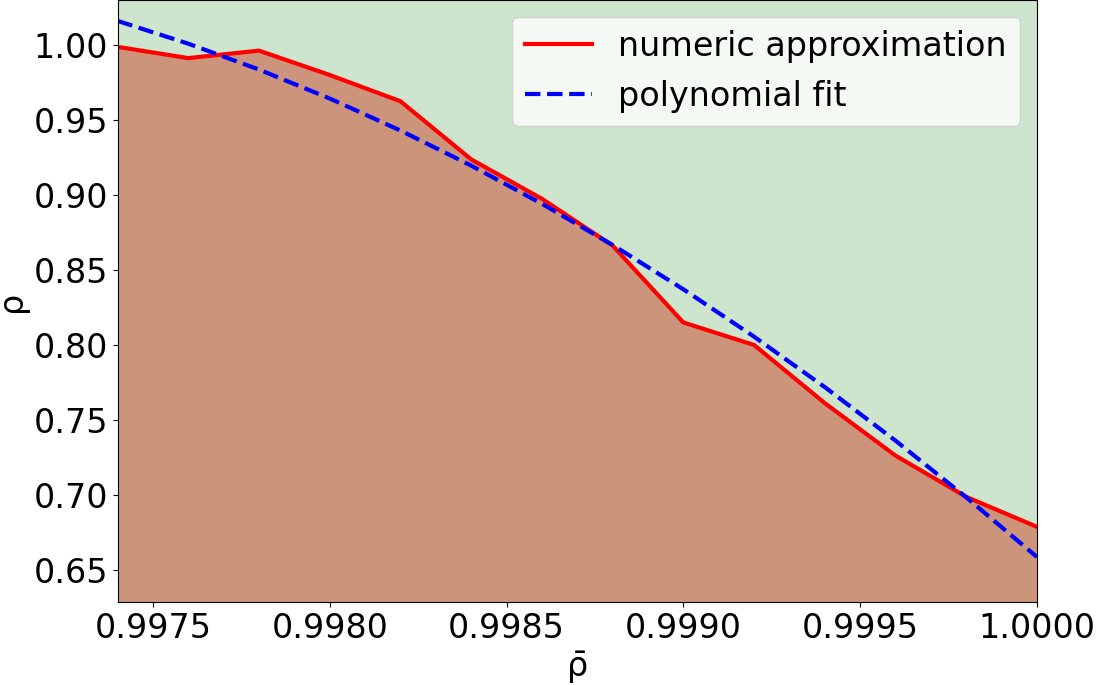}
     \caption{The critical curve separating between the effective (green) and non-effective (red) mitigation regimes when 
     the standard deviation is large $12\sigma = 80[nm]$
     he critical curve has changed compared to the critical curve in \ref{fig: critical line}. The polynomial used for the fitting function is: $a_0 + a_1 \cdot \br + a_1 \cdot \br^2 + a_3 \cdot \br^3$ where  
$a_0 = -10097, a_1 = 5009, a_2 = 20478, a_3 =-15390$.}
     \label{fig: critical correlation function}
\end{figure}

\textit{Generalization: Correlation Curve (Theorem 3)}\textbf{—}In the previous section, we analyzed numerically the critical curve in the $(\rho,\br)$ plane, which indeed was a polynomial function, as can be seen in in Figure \ref{fig: critical line}.
However, in those calculations, the standard deviation of the width error was taken as $1\sigma$ and higher powers could be neglected.
In the following we consider the case where the standard deviation is much larger, $12\sigma = 80[nm]$, so that higher powers cannot be neglected (roughly $ \sim 20\%$ of the waveguides' width). We can see in figure \ref{fig: critical correlation function} that the curve separating the effective from the non-effective mitigation regimes is a polynomial in an higher order compared to the curve shown in Fig. \ref{fig: critical line}. The numerical approximation, seen in Figure \ref{fig: critical correlation function}, is noisy - however, it's very similar to the polynomial estimation function, with maximal difference of only $2\% $, and mean difference smaller than $1\% $.

\newpage
\textit{Fidelity Variance (Theorem 4)}\textbf{—}In previous sections, we presented how the mean fidelity of segmented and uniform DCs changes for different correlation values. In this section, we show how the fidelity standard deviation changes for different correlations, and compare the critical correlation value for the mean fidelity, $\rho_c$, to the critical correlation value for standard deviations, $\rho_c^{std}$.

\begin{figure}[!h]
     \centering
     \includegraphics[width=9cm]{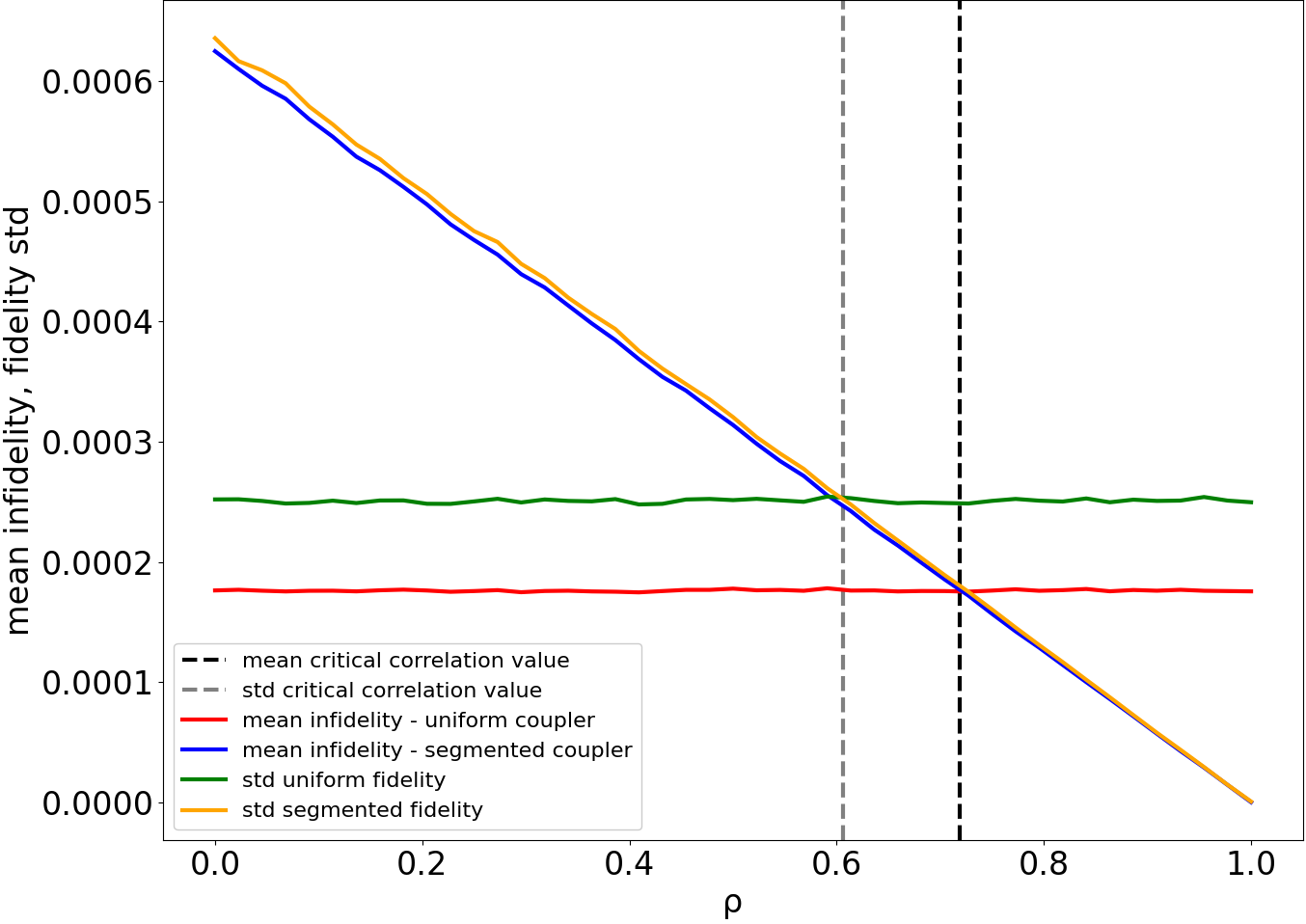}
     \caption{Both mean infidelity and fidelity standard deviation as a function of $\rho$ for a fixed $\sigma$.}
     \label{fig: fidelity vs rho std and mean}
\end{figure}

In Figure \ref{fig: fidelity vs rho std and mean} we plot the fidelity standard deviation as a function of the correlation parameter $\rho$ for
a fixed standard deviation $\sigma$, and compare it to the mean infidelity (which is 1-mean fidelity) as a function of the same correlation parameter $\rho$.
As can be seen from the figure, the fidelity standard deviation shrinks linearly in $\rho$, in the same fashion as the mean fidelity (as can be seen from Fig. \ref{fig: fidelity vs rho}). The dashed gray line in the plot denote $\rho_{c}^{std}$, which is the lowest $\rho$ value where the fidelity standard deviation of the segmented coupler is lower or equal to the fidelity standard deviation of the uniform coupler.

\begin{figure}[!h]
     \centering
     \includegraphics[width=9cm]{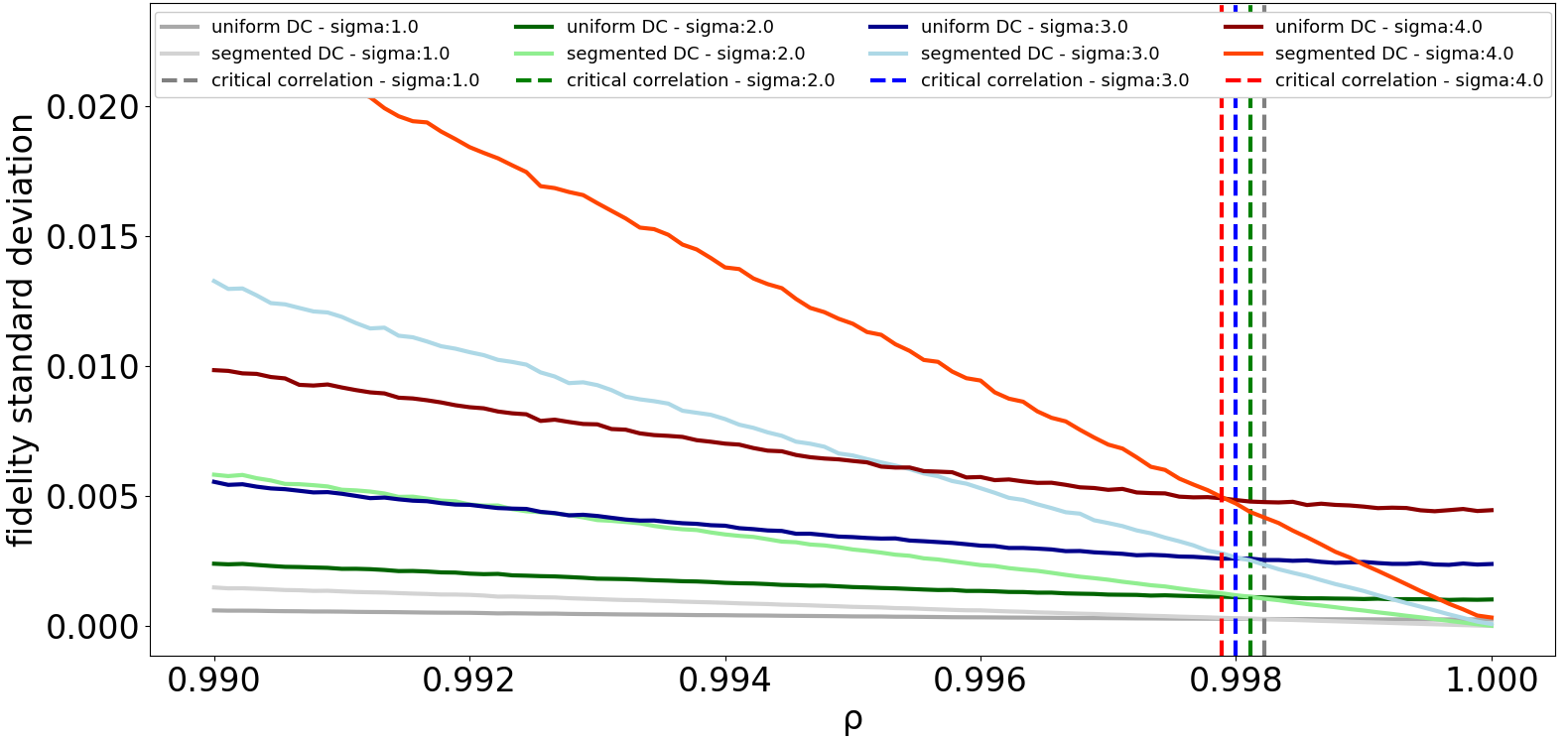}
     \caption{Fidelity standard deviation as a function of $\rho$ for a fixed $\bar{\sigma}$. Here the $\rho=1$ solution applied
     to lower correlation.}
     \label{fig: rho between waveguides vs fidelity std}
\end{figure}

In Figure \ref{fig: rho between waveguides vs fidelity std}
we plot the fidelity standard deviation as a function of $\Tilde{\rho}$ for different values
of the standard deviation $\bar{\sigma}$. As in Figure \ref{fig: rho between waveguides vs fidelity}, the fidelity standard deviation changes in a linear fashion. However, as can be seen from the different dashed lines that indicate the critical parameters $\rho_{c}$, they are dependant on $\sigma$, unlike in Fig. \ref{fig: rho crit between waveguides vs fidelity}.

\begin{figure}[!h]
     \centering
     \includegraphics[width=8cm]{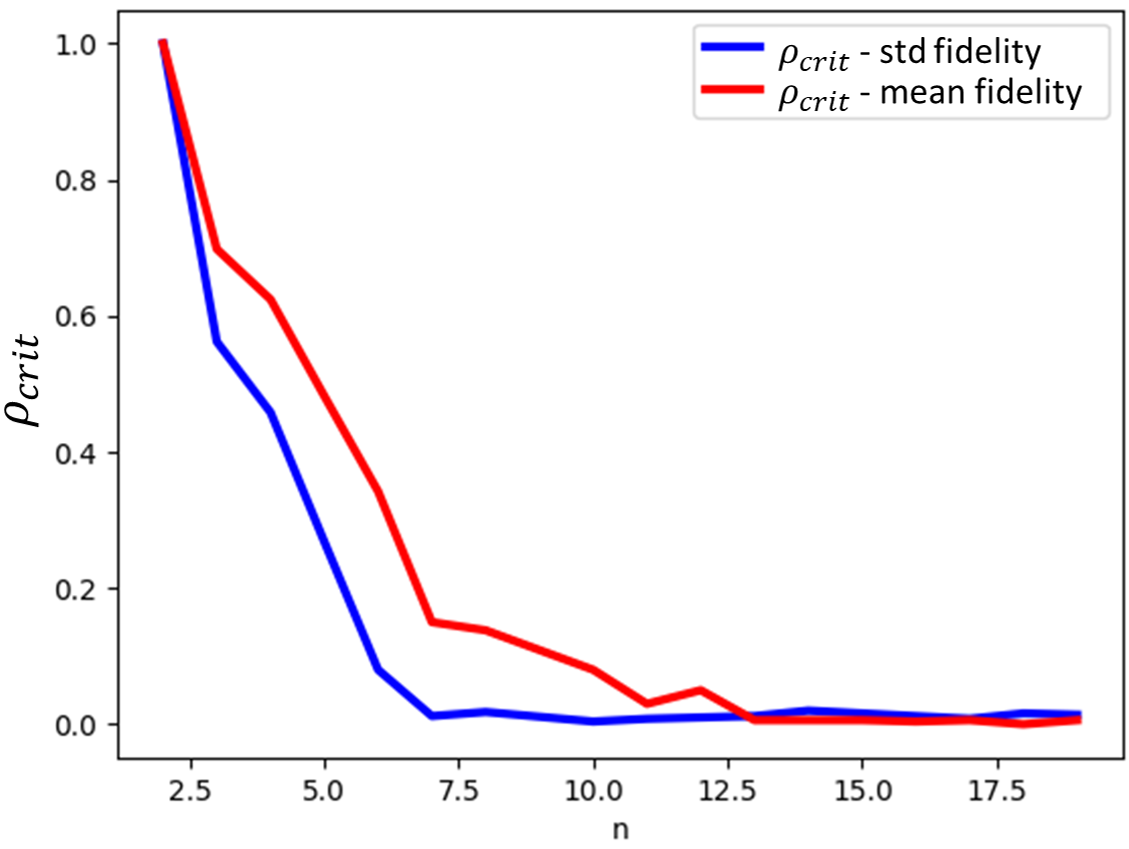}
     \caption{$\rho_c$ and $\rho_c^{std}$ as a function of n, the number of segments in the DC.}
     \label{fig: rho crit std and mean vs n}
\end{figure}

In Figure \ref{fig: rho crit std and mean vs n} we plot the different $\rho_c$ values as a function of the number of segments n. We plot for both the mean infidelity and the fidelity standard deviation. As can be seen from the figure, both $\rho_c$ values decrease as the number of segments increase, as expected. Furthermore, we can see that the rate of descent of $\rho_c^{std}$ is greater than that of $\rho_c$ - this examplifies that for every n, $\rho_c > \rho_c^{std}$.

\newpage

\textit{Conclusion.}\textbf{—} We studied analytically and numerically
the effect of systematic error correlations on the performance 
of the composite pulse segmentation of quantum gates.
We proved that there is a critical correlation line (hypersurface in general) threshold above which the mitigation scheme
decreases the errors. 
The theorems capture the basic effects of the correlations on the 
performance of the CP error mitigation scheme and can bbe generalized in a straightforward way to other covariance matrices, cases where the variance of the error distribution is not small and to other error distributions.
Our work opens a new avenue in the study of CP error mitigation scheme, with important theoretical and experimental implications, not only in the photonic realm, but also in NMR QIPs \cite{Ota_2009}, atomic QIPs \cite{Madjarov_2020} and additional systems.
\vspace{1cm}

\vspace{1cm}

The work of Y.O. is supported in part by  ISF Center of Excellence.

\small
\bibliographystyle{unsrt.bst}
\bibliography{references}

\begin{thebibliography}{10}

\bibitem{Levitt1979}
Malcolm~H. Levitt and Ray Freeman.
\newblock Nmr population inversion using a composite pulse.
\newblock {\em Journal of Magnetic Resonance (1969)}, 33(2):473--476, 1979.

\bibitem{Shaka1985}
A.~J. Shaka.
\newblock Composite pulses for ultra-broadband spin inversion.
\newblock {\em Chemical Physics Letters}, 120(2):201--205, 1985.

\bibitem{Shaka1987}
A.J Shaka and Alexander Pines.
\newblock Symmetric phase-alternating composite pulses.
\newblock {\em Journal of Magnetic Resonance (1969)}, 71(3):495--503, 1987.

\bibitem{Levitt1986}
Malcolm~H. Levitt.
\newblock Composite pulses.
\newblock {\em Progress in Nuclear Magnetic Resonance Spectroscopy},
  18(2):61--122, 1986.

\bibitem{Timoney2008}
Nuala Timoney, V.~Elman, Steffen~J. Glaser, C.~Weiss, M.~Johanning,
  W.~Neuhauser, and Chr. Wunderlich.
\newblock Error-resistant single-qubit gates with trapped ions.
\newblock {\em Physical Review A}, 77(5), 2008.

\bibitem{Torosov2022}
Boyan~T. Torosov and Nikolay~V. Vitanov.
\newblock Experimental demonstration of composite pulses on ibm's quantum
  computer.
\newblock 2022.

\bibitem{partialCorrelated}
James~P. Clemens, Shabnam Siddiqui, and Julio Gea-Banacloche.
\newblock Quantum error correction against correlated noise.
\newblock {\em Phys. Rev. A}, 69:062313, Jun 2004.

\bibitem{nation2021scalable}
Paul~D Nation, Hwajung Kang, Neereja Sundaresan, and Jay~M Gambetta.
\newblock Scalable mitigation of measurement errors on quantum computers.
\newblock {\em PRX Quantum}, 2(4):040326, 2021.

\bibitem{Preskill_1998}
John Preskill.
\newblock Reliable quantum computers.
\newblock {\em Proceedings of the Royal Society of London. Series A:
  Mathematical, Physical and Engineering Sciences}, 454(1969):385--410, jan
  1998.

\bibitem{kyoseva2019detuning}
Elica Kyoseva, Hadar Greener, and Haim Suchowski.
\newblock Detuning-modulated composite pulses for high-fidelity robust quantum
  control.
\newblock {\em Physical Review A}, 100(3):032333, 2019.

\bibitem{YaronIdo}
Ido Kaplan, Muhammad Erew, Yonatan Piasetzky, Moshe Goldstein, Yaron Oz, and
  Haim Suchowski.
\newblock Segmented composite design of robust single-qubit quantum gates.
\newblock 2022.

\bibitem{torosov2022experimental}
Boyan~T. Torosov and Nikolay~V. Vitanov.
\newblock Experimental demonstration of composite pulses on ibm's quantum
  computer, 2022.

\bibitem{O_Brien_2009}
Jeremy~L. O{\textquotesingle}Brien, Akira Furusawa, and Jelena
  Vu{\v{c}}kovi{\'{c}}.
\newblock Photonic quantum technologies.
\newblock {\em Nature Photonics}, 3(12):687--695, dec 2009.

\bibitem{Ding_13}
Yunhong Ding, Jing Xu, Francesco~Da Ros, Bo~Huang, Haiyan Ou, and Christophe
  Peucheret.
\newblock On-chip two-mode division multiplexing using tapered directional
  coupler-based mode multiplexer and demultiplexer.
\newblock {\em Opt. Express}, 21(8):10376--10382, Apr 2013.

\bibitem{Ota_2009}
Yukihiro Ota and Yasushi Kondo.
\newblock Composite pulses in {NMR} as nonadiabatic geometric quantum gates.
\newblock {\em Physical Review A}, 80(2), aug 2009.

\bibitem{Madjarov_2020}
Ivaylo~S. Madjarov, Jacob~P. Covey, Adam~L. Shaw, Joonhee Choi, Anant Kale,
  Alexandre Cooper, Hannes Pichler, Vladimir Schkolnik, Jason~R. Williams, and
  Manuel Endres.
\newblock High-fidelity entanglement and detection of alkaline-earth rydberg
  atoms.
\newblock {\em Nature Physics}, 16(8):857--861, may 2020.

\bibitem{Crespi_2011}
Andrea Crespi, Roberta Ramponi, Roberto Osellame, Linda Sansoni, Irene
  Bongioanni, Fabio Sciarrino, Giuseppe Vallone, and Paolo Mataloni.
\newblock Integrated photonic quantum gates for polarization qubits.
\newblock {\em Nature Communications}, 2(1), nov 2011.

\end{thebibliography}

\appendix

\section{Proof of Theorems}

\subsection{Proof of Theorem 2}
\begin{proof}
We will use the covariance matrix (\ref{cov2}).
 The expectation values of the physical and composite fidelities take the form:
\begin{eqnarray}
\bar{F}_{phy} &=& 1 + b_{phy} \sigma^2 +  c_{phy} \rho \sigma^2   + O(\sigma^4) \ , \nonumber\\
\bar{F}_{cp} &=& 1 + b_{cp}\sigma^2 + c_{cp} \rho \sigma^2 + O(\sigma^4) \ ,
\label{Lin}
\end{eqnarray}
where $b_{phy}= b_{ab}\delta^{ab}, c_{phy}= b_{ab}(H^{ab}-\delta^{ab}), b_{cp} = b_{ab}^{ij} H_{ij}\delta^{ab}, c_{cp} = b_{ab}^{ij} H_{ij}(H^{ab}-\delta^{ab})$.
Equating $\bar{F}_{phy}= \bar{F}_{cp}$ at the critical correlation $\rho_c$ we find:
\begin{equation}
    \rho_c = -\frac{b_{phy}-b_{cp}}{c_{phy}- c_{cp}} \ .
    \label{cor_between_waveguides}
\end{equation}
Note, that unlike (\ref{cor_between_segments}), here $b_{cp}$  and $c_{cp}$ are independent of the number of pulses $n$.
$\bar{F}_{cp}$ is a monotonically increasing function of $\rho$ and is greater
than $\bar{F}_{phy}$ for $\rho > \rho_c$.
\end{proof}
\subsection{Proof of Theorem 3}
\begin{proof}
We will consider the covariance matrix (\ref{covgen}).
 We have:
\begin{eqnarray}
\bar{F}_{phy}  = 1  + (b_{phy} +  c_{phy} \bar{\rho}) \sigma^2 +
O(\sigma^4)  \nonumber\\
\bar{F}_{cp} = 1 + (b_{cp}+ c_{cp}  \bar{\rho} + d_{cp} \rho 
+ e_{cp} \rho \bar{\rho}) \sigma^2 + O(\sigma^4)  \ ,
\end{eqnarray}
where 
\begin{eqnarray}
b_{phy} &=& b_{ab}\delta^{ab},~~~ c_{phy} = b_{ab}(H^{ab}-\delta^{ab})\nonumber\\ 
b_{cp} &=& b_{ab}^{ij}\delta_{ij}\delta^{ab},~~~
c_{cp} = b_{ab}^{ij}\delta_{ij}(H^{ab}-\delta^{ab})\nonumber\\ 
d_{cp} &=& b_{ab}^{ij}\delta^{ab}(H_{ij}-\delta_{ij}),
e_{cp} = b_{ab}^{ij}(H^{ab}-\delta^{ab})(H_{ij}-\delta_{ij})\nonumber\\
\end{eqnarray}

We find that the critical curve ${\cal C}$ is given by:
\begin{eqnarray}
&{\cal C}: E\rho \bar{\rho} + D \rho + C \bar{\rho}  + B  = 0,~~E = e_{cp} \ , \nonumber\\
&D = d_{cp},~C =  c_{cp} - c_{phy}, B = b_{cp}-b_{phy} \ .
\label{corline}
\end{eqnarray}
The region where the composite pulse is effective is the area of the triangle in the 
$(\rho,\bar{\rho})$ plane bounded 
by the curve ${\cal C}$ and the lines $\rho=1$ and $\tilde{\rho}=1$.
\end{proof}

\subsection{Proof of Theorem 4}
\begin{proof}
Consider the covariance matrix of Eq. (\ref{segerrors_ij}) and denote the mean and standard deviations of the fidelities
(\ref{FR}) by $\mu \equiv \mathbf{E}[F], \sigma \equiv \sqrt{\mathbf{E}[F^2] - \mathbf{E}[F]^2}$.
The standard deviation of the physical fidelity (\ref{FR}) takes the form:
\begin{equation}
\sigma_{phy} = \sqrt{2} H^{ab}b_{ab} \sigma^2 + O(\sigma^4)  \ .
\label{sFP}
\end{equation}
Since the mean fidelity is:
\begin{equation}
\mu_{phy} = 1 + H^{ab}b_{ab} \sigma^2 + O(\sigma^4) \ ,
\end{equation}
we have:
\begin{equation}
\frac{\sigma_{phy}}{1-\mu_{phy}} = \sqrt{2} \ ,   
\label{rel1}
\end{equation}
up to $O(\sigma^4)$ corrections.

Similarly, for the CP we have up to $O(\sigma^4)$ corrections:
\begin{equation}
\frac{\sigma_{cp}}{1-\mu_{cp}} = \sqrt{2}G(\rho) \ ,   
\label{rel2}
\end{equation}
where, using $D_{ij} = H_{ij}-\delta_{ij}$, 
\begin{equation}
 G(\rho) = \sqrt{\frac{Tr (b (D+\rho\delta) b (D+\rho \delta)}{(Tr (b (D+\rho\delta)))^2}} \ ,  
\end{equation}
$0 \leq G(\rho) \leq 1, G(1) =1$.
(\ref{rel1}) and (\ref{rel2}) imply that
in the regime where the mean composite pulse fidelity is higher than the mean physical
fidelity:
\begin{equation}
\sigma_{cp} \leq \sqrt{2}(1-\mu_{cp}) \leq \sqrt{2}(1-\mu_{phy}) \leq  \sigma_{phy} \ .   
\end{equation}
\end{proof}

\section{Numerical Simulations}

The main set of parameter values used for the simulations:

\begin{table}[htbp]
\centering
\resizebox{8.5cm}{!}{%
  \begin{tabular}{|c|c|c|}
 \hline
{${w_a}_1,{w_b}_1,z_1$} [$\mu$m] & {${w_a}_2,{w_b}_2,z_2$} [$\mu$m] & {${w_a}_3,{w_b}_3,z_3$} [$\mu$m]\\
 \hline
 {0.4125, 0.4685, 22.5835} &
 {0.449, 0.382, 24.7905} &
 {0.412, 0.468, 22.5195} \\
 \hline
  \end{tabular}}
      \caption{Parameters of the 3-segments pulse used in this section.}
  \label{table: rho_crit vs sigma}
\end{table}

\subsection{Numerical Simulations for Theorem 1 \label{appendix: simulations theorem 1}}

\begin{figure}[!h]
     \centering
     \includegraphics[width=8.5cm]{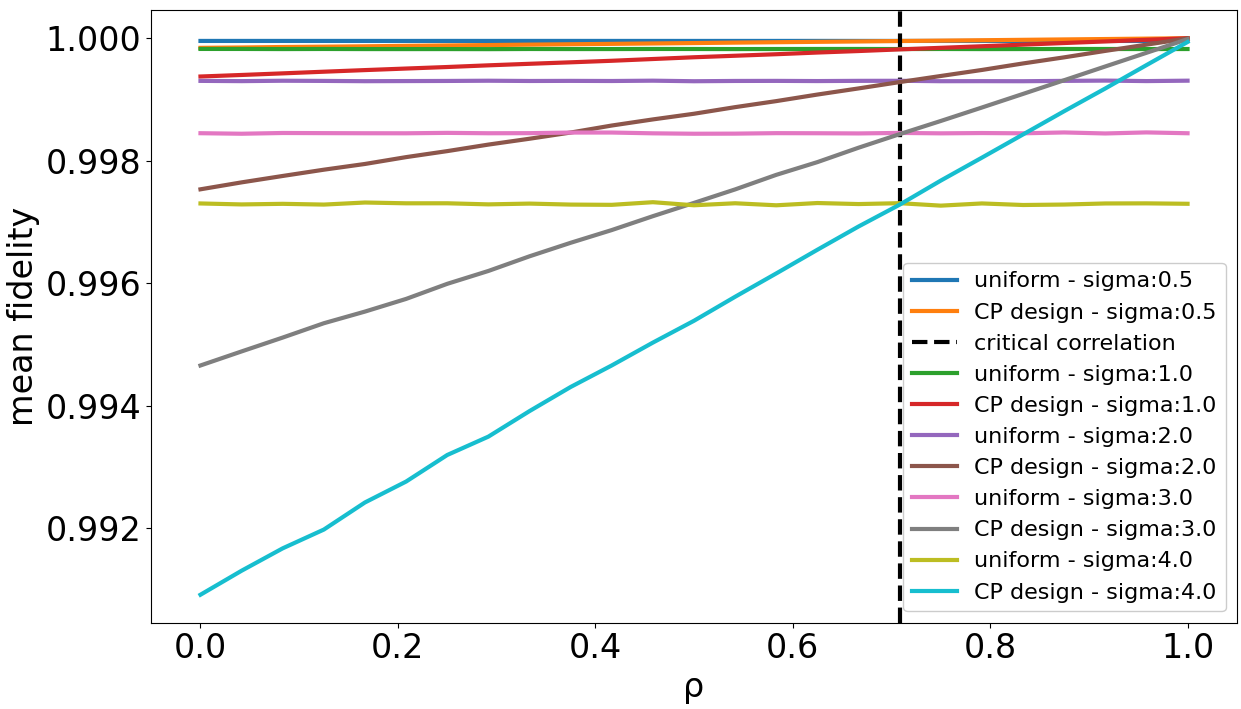}
     \caption{The critical correlation for different values of $\sigma$, where $\rho=1$ is applied to lower correlation. The correlation parameter (\ref{cor_between_segments}) is independent of $\sigma$. The solution is the same for all $\sigma$.}
     \label{fig: rho crit vs sigma}
\end{figure}

In Figure \ref {fig: rho crit vs sigma}
we plot the mean fidelity as a function of $\rho$  for different values
of the standard deviation $\sigma$. As can be seen from the dashed line that indicates
the critical parameter $\rho_{crit}$ it is independent of $\sigma$, as expected from Eq. 
 (\ref{cor_between_segments}).

\newpage

\subsection{Numerical Simulations for Theorem 2
\label{appendix: simulations theorem 2}}

\begin{figure}[!h]
     \centering
     \includegraphics[width=8cm]{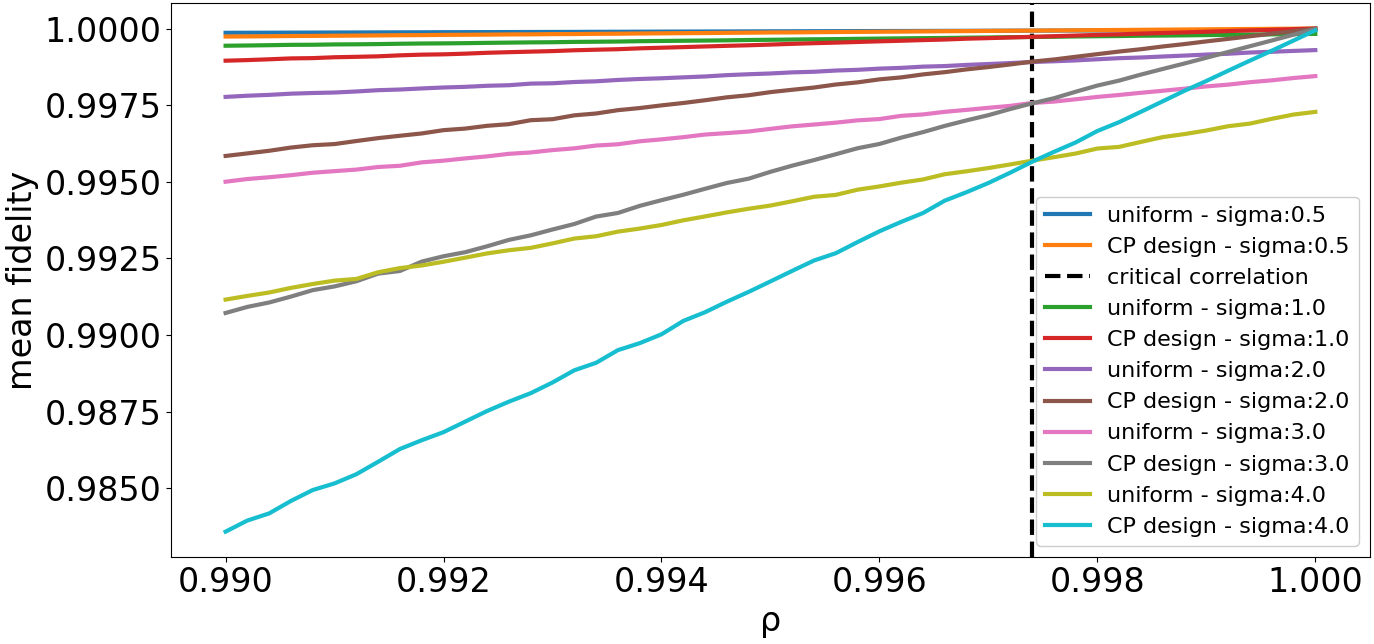}
     \caption{$\rho$ (\ref{cor_between_waveguides}) versus the  fidelity for correlation between waveguides for various $\sigma$ values. We see that $\rho_c$ is independent of the standard deviation.}
     \label{fig: rho crit between waveguides vs fidelity}
\end{figure}

In Figure \ref {fig: rho crit between waveguides vs fidelity}
we plot the mean fidelity as a function of $\rho$  for different values
of the standard deviation $\sigma$. As can be seen from the dashed line that indicates
the critical parameter $\rho_{c}$ it is independent of $\sigma$ (\ref{cor_between_segments}).

\begin{figure}[!h]
     \centering
     \includegraphics[width=8.5cm]{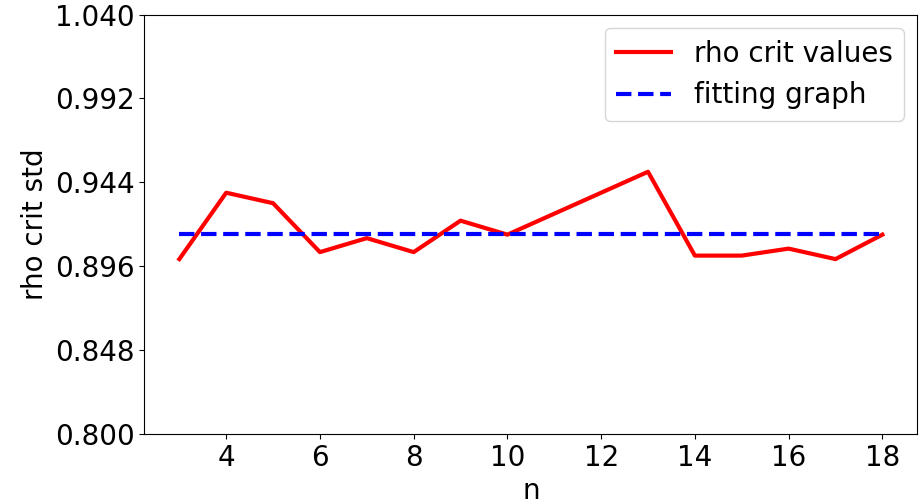}
     \caption{$\rho_{crit}$  (\ref{cor_between_waveguides}) between waveguides as a function of the number of segments $n$. As expected, it is 
     independent of $n$.}
     \label{fig: rho crit between waveguides vs n }
\end{figure}

In Figure \ref{fig: rho crit between waveguides vs n } we show that $\rho_c$ is independent of the number of segments $n$, in contrast to its $\frac{1}{n}$ scaling
in Figure \ref{fig: rho crit vs n}.

\subsection{Fidelity variance numerical Simulations}

\begin{figure}[!h]
     \centering
     \includegraphics[width=9cm]{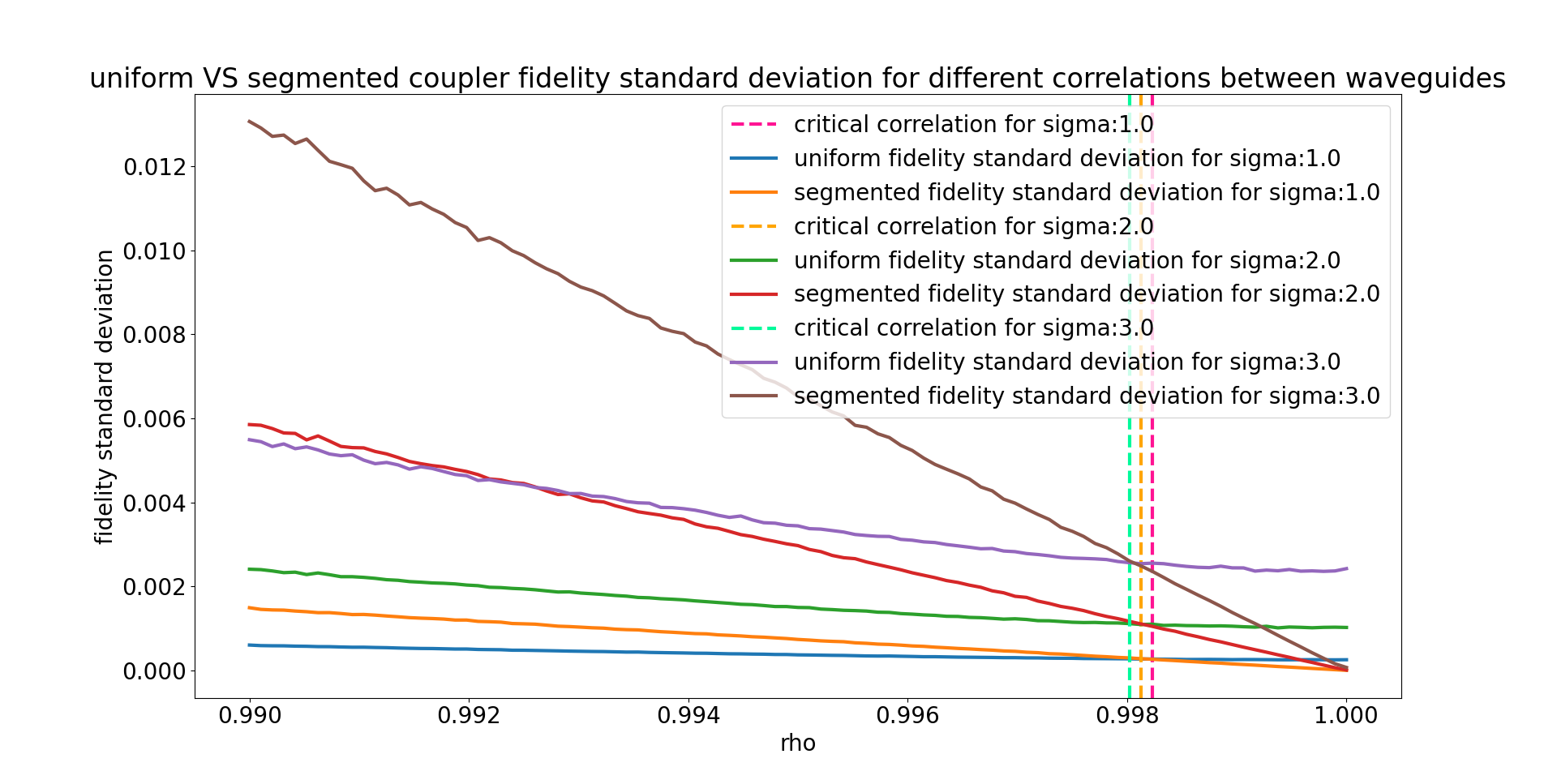}
     \caption{$\rho$ vs fidelity standard deviation for correlation between waveguides for various $\bar{\sigma}$ values. We see that $\rho_c$ is independent of the standard deviation.}
     \label{fig: rho crit between waveguides vs fidelity std}
\end{figure}

In Figure \ref{fig: rho crit between waveguides vs fidelity std}
we plot the fidelity standard deviation as a function of $\rho$  for different values
of the standard deviation $\bar{\sigma}$. As can be seen from the different dashed lines that indicates the critical parameters $\rho_{c}$, the are dependant on $\sigma$, unlike in figure \ref{fig: rho crit between waveguides vs fidelity}.

\begin{figure}[!h]
     \centering
     \includegraphics[width=8cm]{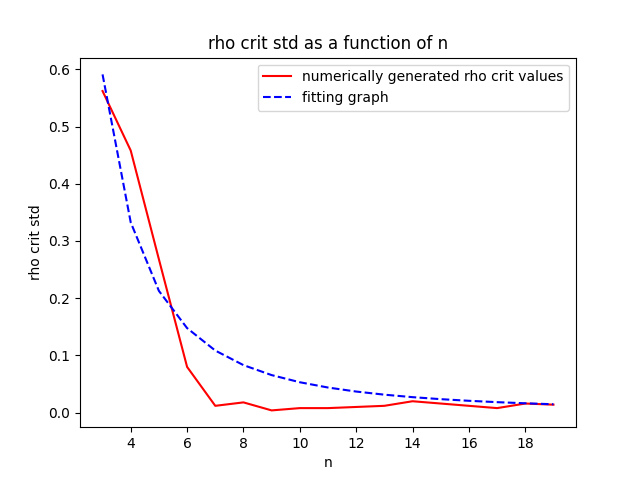}
     \caption{$\rho_{crit}^{std}$ for different number of segments.}
     \label{fig: rho crit std vs n}
\end{figure}

In Figure \ref{fig: rho crit std vs n} we plot the dependence of $\rho_{c}^{std}$ on the number of segments $n$.
The fitting graph used in this figure is:
$$f=\frac{a_1}{n^2}$$
Where $a_1=5.319$.
As can be seen for the mean fidelity graph, additional segments increase the gate robustness at low correlation.

\newpage

\section{Directional Coupler's structure \label{appendix: DC structure}}

Directional couplers (DCs) are often used as quantum gates in photonic QIPs \cite{Crespi_2011}. The DC is a passive optical device which allows the transfer of light between two waveguides. It has a simple structure consisting of two closely coupled parallel waveguides that can be miniaturized to less than several tens of µm [8]. In the coupler, incident light propagating in the waveguide is coupled to the adjacent waveguide through an evanescent field in the gap between the two waveguides. The energy in one waveguide is fully transferred to the other within a distance that is called the complete coupling length. By using different DC lengths, and by adding phase detuning (mainly done by setting different widths for the waveguides within the DC), one can construct different quantum gates, that exhibit weak decoherence. 

\section{All solutions used for approximating $\rho_c$ as a function of the number of segments \label{sets of solutions for rho_crit vs n}}

In order to numerically approximate how $\rho_c$ changes when increasing the number of segments, we required many different solutions: for every number of segments $n$ we required solutions, and for each $n$ we required a set of different solutions - since each solution has a different $\rho_c$ value, we needed many solutions to approximate the minimal $\rho_c$ for the given $n$.

In the tables below, we present all the solutions used in order to approximate the graph presented in Fig \ref{fig: rho crit vs n}.

  \begin{table}[htbp]
  \centering
  \resizebox{8.5cm}{!}{%
    \begin{tabular}{|c|c|c|}
   \hline
  {${w_a}_{0},{w_b}_{0},z_{0}$} [$\mu$m] & {${w_a}_{1},{w_b}_{1},z_{1}$} [$\mu$m] & {${w_a}_{2},{w_b}_{2},z_{2}$} [$\mu$m]\\
   \hline
   {0.451, 0.388, 20.457} &
 {0.385, 0.478, 19.007} &
 {0.49, 0.451, 29.959} \\ 
{0.412, 0.468, 22.583} &
 {0.449, 0.382, 24.791} &
 {0.412, 0.468, 22.519} \\ 

   \hline
    \end{tabular}}
        \caption{Solutions for 3 segments.}
    \label{table: sols for 3 segments.}
  \end{table}

  \begin{table}[htbp]
  \centering
  \resizebox{8.5cm}{!}{%
    \begin{tabular}{|c|c|c|c|}
   \hline
  {${w_a}_{0},{w_b}_{0},z_{0}$} [$\mu$m] & {${w_a}_{1},{w_b}_{1},z_{1}$} [$\mu$m] & {${w_a}_{2},{w_b}_{2},z_{2}$} [$\mu$m] & {${w_a}_{3},{w_b}_{3},z_{3}$} [$\mu$m]\\
   \hline
   {0.408, 0.468, 15.945} &
 {0.351, 0.331, 13.632} &
 {0.482, 0.437, 25.09} &
 {0.311, 0.49, 25.663} \\ 
{0.49, 0.417, 18.97} &
 {0.39, 0.482, 17.592} &
 {0.49, 0.462, 25.799} &
 {0.426, 0.353, 3.459} \\ 
{0.484, 0.49, 31.007} &
 {0.49, 0.375, 12.358} &
 {0.351, 0.49, 17.677} &
 {0.49, 0.31, 6.732} \\ 
{0.445, 0.49, 24.185} &
 {0.49, 0.408, 20.002} &
 {0.423, 0.489, 14.007} &
 {0.439, 0.49, 5.765} \\ 
{0.49, 0.454, 16.881} &
 {0.414, 0.49, 20.573} &
 {0.49, 0.399, 20.814} &
 {0.345, 0.49, 5.804} \\ 
{0.49, 0.402, 13.868} &
 {0.386, 0.49, 19.721} &
 {0.47, 0.368, 9.771} &
 {0.49, 0.483, 21.558} \\ 

   \hline
    \end{tabular}}
        \caption{Solutions for 4 segments.}
    \label{table: sols for 4 segments.}
  \end{table}

  \begin{table}[htbp]
  \centering
  \resizebox{8.5cm}{!}{%
    \begin{tabular}{|c|c|c|c|c|}
   \hline
  {${w_a}_{0},{w_b}_{0},z_{0}$} [$\mu$m] & {${w_a}_{1},{w_b}_{1},z_{1}$} [$\mu$m] & {${w_a}_{2},{w_b}_{2},z_{2}$} [$\mu$m] & {${w_a}_{3},{w_b}_{3},z_{3}$} [$\mu$m] & {${w_a}_{4},{w_b}_{4},z_{4}$} [$\mu$m]\\
   \hline
   {0.443, 0.49, 15.986} &
 {0.38, 0.423, 7.681} &
 {0.49, 0.393, 17.178} &
 {0.421, 0.478, 15.252} &
 {0.445, 0.489, 7.51} \\ 
{0.356, 0.49, 9.232} &
 {0.49, 0.358, 16.822} &
 {0.344, 0.397, 4.505} &
 {0.35, 0.49, 6.46} &
 {0.49, 0.487, 29.384} \\ 
{0.483, 0.49, 28.318} &
 {0.49, 0.379, 12.564} &
 {0.367, 0.49, 19.032} &
 {0.49, 0.326, 3.816} &
 {0.49, 0.328, 3.249} \\ 
{0.47, 0.392, 15.912} &
 {0.367, 0.49, 14.118} &
 {0.434, 0.427, 12.58} &
 {0.49, 0.328, 4.394} &
 {0.49, 0.49, 19.28} \\ 
{0.49, 0.425, 16.062} &
 {0.481, 0.392, 4.765} &
 {0.38, 0.49, 13.357} &
 {0.49, 0.482, 25.258} &
 {0.49, 0.398, 6.643} \\ 
{0.49, 0.487, 17.199} &
 {0.41, 0.49, 12.011} &
 {0.31, 0.353, 2.663} &
 {0.49, 0.383, 17.425} &
 {0.402, 0.485, 13.582} \\ 
{0.487, 0.419, 17.852} &
 {0.376, 0.459, 18.136} &
 {0.353, 0.355, 8.627} &
 {0.49, 0.362, 6.817} &
 {0.49, 0.488, 12.945} \\ 

   \hline
    \end{tabular}}
        \caption{Solutions for 5 segments.}
    \label{table: sols for 5 segments.}
  \end{table}

  \begin{table}[htbp]
  \centering
  \resizebox{16cm}{!}{%
    \begin{tabular}{|c|c|c|c|c|c|}
   \hline
  {${w_a}_{0},{w_b}_{0},z_{0}$} [$\mu$m] & {${w_a}_{1},{w_b}_{1},z_{1}$} [$\mu$m] & {${w_a}_{2},{w_b}_{2},z_{2}$} [$\mu$m] & {${w_a}_{3},{w_b}_{3},z_{3}$} [$\mu$m] & {${w_a}_{4},{w_b}_{4},z_{4}$} [$\mu$m] & {${w_a}_{5},{w_b}_{5},z_{5}$} [$\mu$m]\\
   \hline
   {0.391, 0.352, 9.208} &
 {0.335, 0.49, 6.556} &
 {0.353, 0.36, 4.667} &
 {0.49, 0.488, 13.416} &
 {0.49, 0.397, 16.329} &
 {0.405, 0.49, 14.217} \\ 
{0.375, 0.316, 5.379} &
 {0.362, 0.49, 11.394} &
 {0.49, 0.447, 14.385} &
 {0.49, 0.381, 6.508} &
 {0.49, 0.49, 18.278} &
 {0.4, 0.49, 7.026} \\ 
{0.49, 0.414, 14.105} &
 {0.369, 0.351, 5.672} &
 {0.386, 0.487, 3.171} &
 {0.391, 0.49, 14.173} &
 {0.48, 0.429, 12.761} &
 {0.404, 0.374, 14.934} \\ 
{0.443, 0.357, 14.479} &
 {0.382, 0.439, 13.39} &
 {0.349, 0.432, 12.136} &
 {0.439, 0.437, 13.098} &
 {0.41, 0.337, 15.284} &
 {0.401, 0.425, 12.564} \\ 
{0.445, 0.418, 8.106} &
 {0.465, 0.412, 16.709} &
 {0.388, 0.447, 7.808} &
 {0.385, 0.49, 9.146} &
 {0.351, 0.352, 8.635} &
 {0.49, 0.413, 13.049} \\ 
{0.49, 0.406, 16.361} &
 {0.351, 0.447, 6.991} &
 {0.432, 0.49, 13.359} &
 {0.49, 0.49, 10.548} &
 {0.49, 0.436, 10.347} &
 {0.49, 0.448, 7.306} \\ 

   \hline
    \end{tabular}}
        \caption{Solutions for 6 segments.}
    \label{table: sols for 6 segments.}
  \end{table}

  \begin{table}[htbp]
  \centering
  \resizebox{16cm}{!}{%
    \begin{tabular}{|c|c|c|c|c|c|c|}
   \hline
  {${w_a}_{0},{w_b}_{0},z_{0}$} [$\mu$m] & {${w_a}_{1},{w_b}_{1},z_{1}$} [$\mu$m] & {${w_a}_{2},{w_b}_{2},z_{2}$} [$\mu$m] & {${w_a}_{3},{w_b}_{3},z_{3}$} [$\mu$m] & {${w_a}_{4},{w_b}_{4},z_{4}$} [$\mu$m] & {${w_a}_{5},{w_b}_{5},z_{5}$} [$\mu$m] & {${w_a}_{6},{w_b}_{6},z_{6}$} [$\mu$m]\\
   \hline
   {0.471, 0.348, 5.12} &
 {0.363, 0.397, 11.643} &
 {0.361, 0.472, 7.466} &
 {0.336, 0.338, 11.0} &
 {0.49, 0.376, 10.403} &
 {0.398, 0.386, 9.425} &
 {0.374, 0.49, 5.58} \\ 
{0.49, 0.427, 11.256} &
 {0.49, 0.392, 7.114} &
 {0.366, 0.463, 6.817} &
 {0.397, 0.49, 4.835} &
 {0.385, 0.49, 2.269} &
 {0.49, 0.49, 21.44} &
 {0.49, 0.419, 10.119} \\ 
{0.442, 0.49, 13.929} &
 {0.387, 0.406, 19.68} &
 {0.49, 0.339, 7.457} &
 {0.343, 0.332, 7.022} &
 {0.342, 0.31, 5.078} &
 {0.37, 0.49, 6.399} &
 {0.393, 0.49, 5.029} \\ 
{0.49, 0.43, 8.602} &
 {0.49, 0.39, 5.39} &
 {0.347, 0.36, 7.974} &
 {0.39, 0.49, 13.567} &
 {0.42, 0.393, 12.035} &
 {0.49, 0.411, 6.632} &
 {0.442, 0.424, 8.328} \\ 
{0.49, 0.49, 10.845} &
 {0.49, 0.407, 13.652} &
 {0.394, 0.462, 12.987} &
 {0.389, 0.467, 11.257} &
 {0.484, 0.393, 4.107} &
 {0.49, 0.411, 4.628} &
 {0.49, 0.424, 6.621} \\ 
{0.469, 0.434, 13.611} &
 {0.476, 0.417, 10.946} &
 {0.404, 0.477, 10.819} &
 {0.378, 0.49, 5.836} &
 {0.33, 0.34, 8.32} &
 {0.49, 0.383, 9.072} &
 {0.49, 0.477, 4.633} \\ 
{0.49, 0.441, 11.269} &
 {0.49, 0.427, 1.833} &
 {0.471, 0.424, 11.12} &
 {0.385, 0.431, 6.614} &
 {0.365, 0.49, 8.848} &
 {0.405, 0.388, 14.564} &
 {0.49, 0.419, 11.004} \\ 

   \hline
    \end{tabular}}
        \caption{Solutions for 7 segments.}
    \label{table: sols for 7 segments.}
  \end{table}

  \begin{table}[htbp]
  \centering
  \resizebox{16cm}{!}{%
    \begin{tabular}{|c|c|c|c|c|c|c|c|}
   \hline
  {${w_a}_{0},{w_b}_{0},z_{0}$} [$\mu$m] & {${w_a}_{1},{w_b}_{1},z_{1}$} [$\mu$m] & {${w_a}_{2},{w_b}_{2},z_{2}$} [$\mu$m] & {${w_a}_{3},{w_b}_{3},z_{3}$} [$\mu$m] & {${w_a}_{4},{w_b}_{4},z_{4}$} [$\mu$m] & {${w_a}_{5},{w_b}_{5},z_{5}$} [$\mu$m] & {${w_a}_{6},{w_b}_{6},z_{6}$} [$\mu$m] & {${w_a}_{7},{w_b}_{7},z_{7}$} [$\mu$m]\\
   \hline
   {0.428, 0.38, 11.923} &
 {0.346, 0.49, 9.718} &
 {0.355, 0.342, 16.795} &
 {0.49, 0.354, 7.03} &
 {0.351, 0.334, 8.308} &
 {0.374, 0.423, 3.101} &
 {0.372, 0.49, 5.003} &
 {0.364, 0.49, 0.857} \\ 
{0.49, 0.393, 2.702} &
 {0.49, 0.438, 7.499} &
 {0.474, 0.49, 15.587} &
 {0.393, 0.49, 8.445} &
 {0.384, 0.486, 4.14} &
 {0.468, 0.392, 13.754} &
 {0.49, 0.374, 6.27} &
 {0.336, 0.465, 4.591} \\ 
{0.311, 0.489, 3.638} &
 {0.49, 0.366, 6.216} &
 {0.49, 0.363, 6.233} &
 {0.351, 0.369, 13.589} &
 {0.371, 0.49, 9.118} &
 {0.314, 0.31, 9.213} &
 {0.49, 0.35, 6.295} &
 {0.326, 0.349, 5.303} \\ 
{0.49, 0.49, 8.495} &
 {0.49, 0.381, 10.225} &
 {0.311, 0.324, 4.158} &
 {0.382, 0.49, 11.179} &
 {0.486, 0.49, 5.547} &
 {0.49, 0.49, 11.012} &
 {0.49, 0.424, 3.97} &
 {0.49, 0.433, 7.602} \\ 
{0.386, 0.49, 4.43} &
 {0.383, 0.49, 4.406} &
 {0.381, 0.393, 9.533} &
 {0.49, 0.362, 9.302} &
 {0.49, 0.483, 13.017} &
 {0.446, 0.49, 15.374} &
 {0.411, 0.49, 2.659} &
 {0.49, 0.482, 5.839} \\ 

   \hline
    \end{tabular}}
        \caption{Solutions for 8 segments.}
    \label{table: sols for 8 segments.}
  \end{table}

  \begin{table}[htbp]
  \centering
  \resizebox{16cm}{!}{%
    \begin{tabular}{|c|c|c|c|c|c|c|c|c|}
   \hline
  {${w_a}_{0},{w_b}_{0},z_{0}$} [$\mu$m] & {${w_a}_{1},{w_b}_{1},z_{1}$} [$\mu$m] & {${w_a}_{2},{w_b}_{2},z_{2}$} [$\mu$m] & {${w_a}_{3},{w_b}_{3},z_{3}$} [$\mu$m] & {${w_a}_{4},{w_b}_{4},z_{4}$} [$\mu$m] & {${w_a}_{5},{w_b}_{5},z_{5}$} [$\mu$m] & {${w_a}_{6},{w_b}_{6},z_{6}$} [$\mu$m] & {${w_a}_{7},{w_b}_{7},z_{7}$} [$\mu$m] & {${w_a}_{8},{w_b}_{8},z_{8}$} [$\mu$m]\\
   \hline
   {0.49, 0.404, 11.196} &
 {0.326, 0.337, 6.149} &
 {0.392, 0.482, 9.422} &
 {0.421, 0.489, 6.384} &
 {0.31, 0.326, 3.558} &
 {0.49, 0.395, 3.825} &
 {0.49, 0.43, 8.653} &
 {0.49, 0.461, 4.654} &
 {0.49, 0.49, 5.755} \\ 
{0.49, 0.396, 10.93} &
 {0.31, 0.31, 3.087} &
 {0.366, 0.49, 9.726} &
 {0.375, 0.31, 2.485} &
 {0.353, 0.49, 5.15} &
 {0.465, 0.394, 1.217} &
 {0.478, 0.39, 4.822} &
 {0.49, 0.403, 6.818} &
 {0.49, 0.49, 16.423} \\ 
{0.351, 0.321, 8.649} &
 {0.344, 0.49, 7.514} &
 {0.35, 0.316, 4.155} &
 {0.334, 0.362, 8.089} &
 {0.49, 0.366, 6.25} &
 {0.49, 0.382, 2.889} &
 {0.477, 0.382, 2.525} &
 {0.313, 0.31, 8.201} &
 {0.392, 0.49, 8.78} \\ 
{0.389, 0.465, 11.302} &
 {0.49, 0.329, 0.324} &
 {0.49, 0.352, 10.43} &
 {0.31, 0.353, 4.388} &
 {0.49, 0.49, 15.845} &
 {0.38, 0.49, 3.959} &
 {0.378, 0.49, 7.062} &
 {0.439, 0.401, 5.304} &
 {0.49, 0.367, 5.359} \\ 
{0.431, 0.488, 6.449} &
 {0.382, 0.49, 8.1} &
 {0.342, 0.31, 7.307} &
 {0.49, 0.369, 5.235} &
 {0.374, 0.38, 6.769} &
 {0.482, 0.4, 6.161} &
 {0.35, 0.31, 4.93} &
 {0.371, 0.49, 4.473} &
 {0.437, 0.49, 12.887} \\ 
{0.412, 0.481, 5.708} &
 {0.399, 0.475, 3.465} &
 {0.432, 0.463, 6.019} &
 {0.44, 0.394, 7.461} &
 {0.49, 0.395, 10.987} &
 {0.358, 0.368, 8.043} &
 {0.392, 0.49, 7.055} &
 {0.406, 0.49, 6.886} &
 {0.424, 0.353, 5.659} \\ 
{0.35, 0.413, 8.046} &
 {0.49, 0.398, 10.895} &
 {0.487, 0.406, 6.317} &
 {0.31, 0.31, 4.276} &
 {0.397, 0.49, 11.158} &
 {0.372, 0.49, 2.642} &
 {0.363, 0.346, 6.379} &
 {0.348, 0.364, 4.451} &
 {0.49, 0.368, 4.78} \\ 

   \hline
    \end{tabular}}
        \caption{Solutions for 9 segments.}
    \label{table: sols for 9 segments.}
  \end{table}

  \begin{table}[htbp]
  \centering
  \resizebox{16cm}{!}{%
    \begin{tabular}{|c|c|c|c|c|c|c|c|c|c|}
   \hline
  {${w_a}_{0},{w_b}_{0},z_{0}$} [$\mu$m] & {${w_a}_{1},{w_b}_{1},z_{1}$} [$\mu$m] & {${w_a}_{2},{w_b}_{2},z_{2}$} [$\mu$m] & {${w_a}_{3},{w_b}_{3},z_{3}$} [$\mu$m] & {${w_a}_{4},{w_b}_{4},z_{4}$} [$\mu$m] & {${w_a}_{5},{w_b}_{5},z_{5}$} [$\mu$m] & {${w_a}_{6},{w_b}_{6},z_{6}$} [$\mu$m] & {${w_a}_{7},{w_b}_{7},z_{7}$} [$\mu$m] & {${w_a}_{8},{w_b}_{8},z_{8}$} [$\mu$m] & {${w_a}_{9},{w_b}_{9},z_{9}$} [$\mu$m]\\
   \hline
   {0.489, 0.382, 9.228} &
 {0.365, 0.456, 5.029} &
 {0.378, 0.49, 1.946} &
 {0.425, 0.49, 10.121} &
 {0.49, 0.49, 14.557} &
 {0.31, 0.385, 0.743} &
 {0.49, 0.429, 4.573} &
 {0.49, 0.368, 8.767} &
 {0.354, 0.448, 3.747} &
 {0.373, 0.49, 3.456} \\ 
{0.395, 0.49, 8.879} &
 {0.422, 0.384, 9.514} &
 {0.49, 0.4, 5.582} &
 {0.49, 0.397, 6.126} &
 {0.31, 0.367, 2.336} &
 {0.49, 0.484, 6.356} &
 {0.425, 0.483, 4.298} &
 {0.383, 0.49, 8.989} &
 {0.358, 0.323, 5.783} &
 {0.48, 0.361, 1.946} \\ 
{0.49, 0.396, 5.278} &
 {0.49, 0.379, 4.402} &
 {0.36, 0.371, 7.882} &
 {0.375, 0.414, 4.133} &
 {0.386, 0.49, 8.946} &
 {0.37, 0.424, 4.688} &
 {0.424, 0.389, 4.206} &
 {0.426, 0.386, 8.229} &
 {0.49, 0.389, 5.763} &
 {0.391, 0.397, 8.525} \\ 
{0.358, 0.49, 5.416} &
 {0.334, 0.328, 4.726} &
 {0.39, 0.365, 11.02} &
 {0.49, 0.373, 4.908} &
 {0.49, 0.372, 2.042} &
 {0.373, 0.354, 7.824} &
 {0.376, 0.413, 4.576} &
 {0.378, 0.456, 2.111} &
 {0.371, 0.49, 9.776} &
 {0.396, 0.334, 8.179} \\ 
{0.362, 0.31, 6.516} &
 {0.367, 0.49, 7.656} &
 {0.358, 0.49, 2.766} &
 {0.352, 0.489, 0.15} &
 {0.352, 0.338, 9.277} &
 {0.49, 0.485, 4.832} &
 {0.49, 0.371, 8.913} &
 {0.31, 0.386, 1.497} &
 {0.323, 0.31, 10.209} &
 {0.363, 0.49, 6.314} \\ 
{0.402, 0.362, 9.459} &
 {0.376, 0.49, 1.987} &
 {0.373, 0.49, 0.858} &
 {0.377, 0.49, 7.696} &
 {0.34, 0.344, 9.416} &
 {0.481, 0.381, 6.402} &
 {0.49, 0.391, 7.23} &
 {0.323, 0.353, 4.207} &
 {0.378, 0.31, 1.867} &
 {0.402, 0.488, 9.616} \\ 
{0.49, 0.423, 9.771} &
 {0.49, 0.387, 3.69} &
 {0.31, 0.318, 5.692} &
 {0.378, 0.49, 5.395} &
 {0.384, 0.467, 4.71} &
 {0.391, 0.49, 2.326} &
 {0.354, 0.363, 8.947} &
 {0.49, 0.391, 2.637} &
 {0.481, 0.405, 8.503} &
 {0.49, 0.49, 8.533} \\ 
{0.471, 0.489, 2.21} &
 {0.386, 0.432, 5.37} &
 {0.45, 0.439, 5.216} &
 {0.49, 0.379, 4.673} &
 {0.49, 0.397, 5.86} &
 {0.328, 0.325, 8.972} &
 {0.374, 0.49, 9.957} &
 {0.359, 0.375, 10.624} &
 {0.476, 0.373, 3.136} &
 {0.488, 0.374, 4.398} \\ 
{0.31, 0.378, 4.085} &
 {0.49, 0.393, 5.795} &
 {0.49, 0.384, 7.062} &
 {0.31, 0.315, 7.04} &
 {0.389, 0.49, 7.954} &
 {0.412, 0.49, 4.531} &
 {0.335, 0.331, 7.945} &
 {0.31, 0.361, 2.623} &
 {0.49, 0.387, 4.184} &
 {0.49, 0.416, 5.483} \\ 

   \hline
    \end{tabular}}
        \caption{Solutions for 10 segments.}
    \label{table: sols for 10 segments.}
  \end{table}

  \begin{table}[htbp]
  \centering
  \resizebox{16cm}{!}{%
    \begin{tabular}{|c|c|c|c|c|c|c|c|c|c|c|}
   \hline
  {${w_a}_{0},{w_b}_{0},z_{0}$} [$\mu$m] & {${w_a}_{1},{w_b}_{1},z_{1}$} [$\mu$m] & {${w_a}_{2},{w_b}_{2},z_{2}$} [$\mu$m] & {${w_a}_{3},{w_b}_{3},z_{3}$} [$\mu$m] & {${w_a}_{4},{w_b}_{4},z_{4}$} [$\mu$m] & {${w_a}_{5},{w_b}_{5},z_{5}$} [$\mu$m] & {${w_a}_{6},{w_b}_{6},z_{6}$} [$\mu$m] & {${w_a}_{7},{w_b}_{7},z_{7}$} [$\mu$m] & {${w_a}_{8},{w_b}_{8},z_{8}$} [$\mu$m] & {${w_a}_{9},{w_b}_{9},z_{9}$} [$\mu$m] & {${w_a}_{10},{w_b}_{10},z_{10}$} [$\mu$m]\\
   \hline
   {0.49, 0.49, 4.192} &
 {0.399, 0.49, 6.57} &
 {0.374, 0.49, 4.024} &
 {0.333, 0.315, 8.541} &
 {0.49, 0.381, 4.066} &
 {0.49, 0.384, 2.421} &
 {0.49, 0.454, 6.887} &
 {0.49, 0.49, 5.426} &
 {0.356, 0.31, 5.39} &
 {0.381, 0.49, 7.069} &
 {0.444, 0.49, 5.795} \\ 
{0.49, 0.339, 3.65} &
 {0.351, 0.334, 7.89} &
 {0.35, 0.49, 1.531} &
 {0.359, 0.49, 9.808} &
 {0.31, 0.31, 6.519} &
 {0.488, 0.368, 3.096} &
 {0.49, 0.37, 5.464} &
 {0.436, 0.369, 5.729} &
 {0.31, 0.445, 1.783} &
 {0.31, 0.468, 0.941} &
 {0.48, 0.49, 14.284} \\ 
{0.439, 0.358, 7.518} &
 {0.381, 0.49, 8.486} &
 {0.382, 0.484, 6.301} &
 {0.397, 0.377, 7.361} &
 {0.49, 0.403, 3.327} &
 {0.49, 0.403, 1.997} &
 {0.49, 0.403, 3.024} &
 {0.49, 0.405, 3.38} &
 {0.49, 0.478, 9.688} &
 {0.397, 0.377, 2.509} &
 {0.411, 0.49, 8.156} \\ 
{0.363, 0.413, 9.961} &
 {0.49, 0.374, 4.979} &
 {0.49, 0.367, 6.869} &
 {0.321, 0.339, 8.096} &
 {0.355, 0.31, 2.352} &
 {0.364, 0.49, 6.131} &
 {0.372, 0.49, 1.405} &
 {0.364, 0.418, 5.877} &
 {0.31, 0.31, 2.743} &
 {0.398, 0.366, 6.428} &
 {0.49, 0.383, 4.767} \\ 

   \hline
    \end{tabular}}
        \caption{Solutions for 11 segments.}
    \label{table: sols for 11 segments.}
  \end{table}

  \begin{table}[htbp]
  \centering
  \resizebox{16cm}{!}{%
    \begin{tabular}{|c|c|c|c|c|c|c|c|c|c|c|c|}
   \hline
  {${w_a}_{0},{w_b}_{0},z_{0}$} [$\mu$m] & {${w_a}_{1},{w_b}_{1},z_{1}$} [$\mu$m] & {${w_a}_{2},{w_b}_{2},z_{2}$} [$\mu$m] & {${w_a}_{3},{w_b}_{3},z_{3}$} [$\mu$m] & {${w_a}_{4},{w_b}_{4},z_{4}$} [$\mu$m] & {${w_a}_{5},{w_b}_{5},z_{5}$} [$\mu$m] & {${w_a}_{6},{w_b}_{6},z_{6}$} [$\mu$m] & {${w_a}_{7},{w_b}_{7},z_{7}$} [$\mu$m] & {${w_a}_{8},{w_b}_{8},z_{8}$} [$\mu$m] & {${w_a}_{9},{w_b}_{9},z_{9}$} [$\mu$m] & {${w_a}_{10},{w_b}_{10},z_{10}$} [$\mu$m] & {${w_a}_{11},{w_b}_{11},z_{11}$} [$\mu$m]\\
   \hline
   {0.49, 0.403, 1.083} &
 {0.49, 0.4, 1.052} &
 {0.49, 0.42, 7.177} &
 {0.49, 0.49, 4.411} &
 {0.345, 0.33, 7.733} &
 {0.368, 0.49, 8.192} &
 {0.369, 0.415, 7.03} &
 {0.473, 0.417, 5.61} &
 {0.49, 0.487, 5.814} &
 {0.49, 0.362, 8.548} &
 {0.316, 0.373, 3.615} &
 {0.337, 0.479, 1.288} \\ 
{0.49, 0.425, 9.422} &
 {0.49, 0.384, 3.164} &
 {0.49, 0.373, 0.747} &
 {0.49, 0.374, 2.952} &
 {0.332, 0.399, 9.275} &
 {0.383, 0.429, 6.068} &
 {0.46, 0.49, 5.594} &
 {0.444, 0.478, 7.107} &
 {0.314, 0.342, 7.184} &
 {0.49, 0.351, 2.644} &
 {0.49, 0.364, 6.772} &
 {0.31, 0.362, 2.263} \\ 
{0.351, 0.379, 11.747} &
 {0.49, 0.352, 5.578} &
 {0.49, 0.355, 2.4} &
 {0.364, 0.349, 5.504} &
 {0.347, 0.392, 3.382} &
 {0.355, 0.31, 3.214} &
 {0.362, 0.49, 4.404} &
 {0.373, 0.49, 7.06} &
 {0.315, 0.31, 2.643} &
 {0.31, 0.31, 0.516} &
 {0.393, 0.389, 4.313} &
 {0.49, 0.4, 9.089} \\ 

   \hline
    \end{tabular}}
        \caption{Solutions for 12 segments.}
    \label{table: sols for 12 segments.}
  \end{table}

  \begin{table}[htbp]
  \centering
  \resizebox{16cm}{!}{%
    \begin{tabular}{|c|c|c|c|c|c|c|c|c|c|c|c|c|}
   \hline
  {${w_a}_{0},{w_b}_{0},z_{0}$} [$\mu$m] & {${w_a}_{1},{w_b}_{1},z_{1}$} [$\mu$m] & {${w_a}_{2},{w_b}_{2},z_{2}$} [$\mu$m] & {${w_a}_{3},{w_b}_{3},z_{3}$} [$\mu$m] & {${w_a}_{4},{w_b}_{4},z_{4}$} [$\mu$m] & {${w_a}_{5},{w_b}_{5},z_{5}$} [$\mu$m] & {${w_a}_{6},{w_b}_{6},z_{6}$} [$\mu$m] & {${w_a}_{7},{w_b}_{7},z_{7}$} [$\mu$m] & {${w_a}_{8},{w_b}_{8},z_{8}$} [$\mu$m] & {${w_a}_{9},{w_b}_{9},z_{9}$} [$\mu$m] & {${w_a}_{10},{w_b}_{10},z_{10}$} [$\mu$m] & {${w_a}_{11},{w_b}_{11},z_{11}$} [$\mu$m] & {${w_a}_{12},{w_b}_{12},z_{12}$} [$\mu$m]\\
   \hline
   {0.414, 0.49, 2.117} &
 {0.369, 0.424, 2.762} &
 {0.381, 0.318, 0.718} &
 {0.378, 0.469, 6.632} &
 {0.41, 0.422, 8.138} &
 {0.49, 0.367, 4.481} &
 {0.462, 0.379, 7.236} &
 {0.484, 0.398, 4.853} &
 {0.381, 0.405, 6.598} &
 {0.403, 0.488, 6.679} &
 {0.389, 0.441, 7.702} &
 {0.49, 0.49, 5.927} &
 {0.465, 0.318, 0.695} \\ 
{0.386, 0.49, 6.431} &
 {0.329, 0.339, 4.871} &
 {0.49, 0.387, 5.035} &
 {0.49, 0.396, 4.356} &
 {0.49, 0.49, 3.639} &
 {0.49, 0.392, 4.817} &
 {0.345, 0.349, 6.746} &
 {0.381, 0.49, 3.357} &
 {0.381, 0.49, 2.777} &
 {0.369, 0.49, 4.984} &
 {0.348, 0.31, 3.242} &
 {0.333, 0.372, 1.921} &
 {0.436, 0.368, 6.548} \\ 
{0.49, 0.354, 2.518} &
 {0.354, 0.334, 8.27} &
 {0.37, 0.484, 4.347} &
 {0.376, 0.49, 2.824} &
 {0.373, 0.49, 2.746} &
 {0.366, 0.392, 5.349} &
 {0.376, 0.344, 6.638} &
 {0.31, 0.357, 2.299} &
 {0.49, 0.366, 5.375} &
 {0.49, 0.372, 4.238} &
 {0.31, 0.31, 7.572} &
 {0.368, 0.49, 1.871} &
 {0.375, 0.49, 3.704} \\ 
{0.49, 0.417, 3.389} &
 {0.484, 0.394, 7.267} &
 {0.31, 0.367, 1.983} &
 {0.351, 0.34, 6.131} &
 {0.384, 0.49, 5.431} &
 {0.385, 0.49, 3.775} &
 {0.387, 0.49, 1.255} &
 {0.379, 0.43, 4.106} &
 {0.369, 0.358, 4.577} &
 {0.41, 0.374, 6.789} &
 {0.49, 0.374, 7.615} &
 {0.31, 0.355, 3.912} &
 {0.41, 0.444, 2.98} \\ 
{0.397, 0.384, 5.181} &
 {0.434, 0.476, 5.215} &
 {0.397, 0.479, 5.742} &
 {0.377, 0.443, 5.063} &
 {0.389, 0.395, 7.108} &
 {0.489, 0.406, 4.83} &
 {0.485, 0.365, 6.307} &
 {0.431, 0.407, 0.481} &
 {0.479, 0.362, 4.323} &
 {0.359, 0.41, 6.818} &
 {0.375, 0.428, 7.504} &
 {0.403, 0.42, 3.18} &
 {0.39, 0.479, 3.184} \\ 

   \hline
    \end{tabular}}
        \caption{Solutions for 13 segments.}
    \label{table: sols for 13 segments.}
  \end{table}

  \begin{table}[htbp]
  \centering
  \resizebox{16cm}{!}{%
    \begin{tabular}{|c|c|c|c|c|c|c|c|c|c|c|c|c|c|}
   \hline
  {${w_a}_{0},{w_b}_{0},z_{0}$} [$\mu$m] & {${w_a}_{1},{w_b}_{1},z_{1}$} [$\mu$m] & {${w_a}_{2},{w_b}_{2},z_{2}$} [$\mu$m] & {${w_a}_{3},{w_b}_{3},z_{3}$} [$\mu$m] & {${w_a}_{4},{w_b}_{4},z_{4}$} [$\mu$m] & {${w_a}_{5},{w_b}_{5},z_{5}$} [$\mu$m] & {${w_a}_{6},{w_b}_{6},z_{6}$} [$\mu$m] & {${w_a}_{7},{w_b}_{7},z_{7}$} [$\mu$m] & {${w_a}_{8},{w_b}_{8},z_{8}$} [$\mu$m] & {${w_a}_{9},{w_b}_{9},z_{9}$} [$\mu$m] & {${w_a}_{10},{w_b}_{10},z_{10}$} [$\mu$m] & {${w_a}_{11},{w_b}_{11},z_{11}$} [$\mu$m] & {${w_a}_{12},{w_b}_{12},z_{12}$} [$\mu$m] & {${w_a}_{13},{w_b}_{13},z_{13}$} [$\mu$m]\\
   \hline
   {0.49, 0.462, 3.916} &
 {0.449, 0.35, 8.388} &
 {0.366, 0.484, 7.959} &
 {0.404, 0.49, 6.119} &
 {0.382, 0.45, 7.201} &
 {0.482, 0.371, 6.152} &
 {0.475, 0.356, 0.66} &
 {0.382, 0.373, 5.584} &
 {0.316, 0.37, 1.035} &
 {0.444, 0.355, 1.144} &
 {0.49, 0.355, 4.821} &
 {0.449, 0.354, 0.357} &
 {0.381, 0.371, 4.849} &
 {0.388, 0.49, 6.662} \\ 
{0.49, 0.472, 4.515} &
 {0.4, 0.49, 6.919} &
 {0.39, 0.31, 1.035} &
 {0.385, 0.49, 2.803} &
 {0.49, 0.49, 6.658} &
 {0.332, 0.354, 6.881} &
 {0.49, 0.359, 2.935} &
 {0.49, 0.372, 6.102} &
 {0.4, 0.358, 6.711} &
 {0.405, 0.456, 4.023} &
 {0.421, 0.49, 6.067} &
 {0.417, 0.49, 2.374} &
 {0.417, 0.49, 1.713} &
 {0.427, 0.49, 3.528} \\ 
{0.397, 0.49, 6.267} &
 {0.375, 0.49, 2.79} &
 {0.347, 0.31, 4.02} &
 {0.31, 0.373, 1.71} &
 {0.454, 0.385, 6.21} &
 {0.49, 0.4, 5.897} &
 {0.47, 0.4, 7.432} &
 {0.372, 0.471, 4.644} &
 {0.377, 0.31, 2.572} &
 {0.379, 0.49, 3.92} &
 {0.387, 0.49, 1.551} &
 {0.389, 0.49, 3.726} &
 {0.489, 0.49, 6.251} &
 {0.407, 0.346, 2.931} \\ 
{0.49, 0.49, 7.323} &
 {0.49, 0.465, 1.408} &
 {0.49, 0.415, 1.795} &
 {0.49, 0.391, 3.292} &
 {0.49, 0.374, 1.639} &
 {0.31, 0.387, 2.241} &
 {0.49, 0.411, 5.299} &
 {0.332, 0.31, 5.862} &
 {0.359, 0.49, 3.859} &
 {0.373, 0.49, 5.971} &
 {0.362, 0.391, 8.0} &
 {0.475, 0.367, 3.729} &
 {0.31, 0.364, 3.189} &
 {0.49, 0.366, 6.382} \\ 

   \hline
    \end{tabular}}
        \caption{Solutions for 14 segments.}
    \label{table: sols for 14 segments.}
  \end{table}

  \begin{table}[htbp]
  \centering
  \resizebox{16cm}{!}{%
    \begin{tabular}{|c|c|c|c|c|c|c|c|c|c|c|c|c|c|c|}
   \hline
  {${w_a}_{0},{w_b}_{0},z_{0}$} [$\mu$m] & {${w_a}_{1},{w_b}_{1},z_{1}$} [$\mu$m] & {${w_a}_{2},{w_b}_{2},z_{2}$} [$\mu$m] & {${w_a}_{3},{w_b}_{3},z_{3}$} [$\mu$m] & {${w_a}_{4},{w_b}_{4},z_{4}$} [$\mu$m] & {${w_a}_{5},{w_b}_{5},z_{5}$} [$\mu$m] & {${w_a}_{6},{w_b}_{6},z_{6}$} [$\mu$m] & {${w_a}_{7},{w_b}_{7},z_{7}$} [$\mu$m] & {${w_a}_{8},{w_b}_{8},z_{8}$} [$\mu$m] & {${w_a}_{9},{w_b}_{9},z_{9}$} [$\mu$m] & {${w_a}_{10},{w_b}_{10},z_{10}$} [$\mu$m] & {${w_a}_{11},{w_b}_{11},z_{11}$} [$\mu$m] & {${w_a}_{12},{w_b}_{12},z_{12}$} [$\mu$m] & {${w_a}_{13},{w_b}_{13},z_{13}$} [$\mu$m] & {${w_a}_{14},{w_b}_{14},z_{14}$} [$\mu$m]\\
   \hline
   {0.371, 0.49, 5.303} &
 {0.31, 0.31, 3.319} &
 {0.445, 0.371, 7.004} &
 {0.31, 0.379, 1.399} &
 {0.49, 0.375, 0.999} &
 {0.49, 0.377, 4.674} &
 {0.49, 0.369, 2.606} &
 {0.31, 0.312, 6.281} &
 {0.372, 0.471, 5.262} &
 {0.374, 0.49, 4.734} &
 {0.373, 0.31, 1.101} &
 {0.359, 0.49, 2.407} &
 {0.346, 0.319, 4.865} &
 {0.49, 0.486, 4.314} &
 {0.49, 0.379, 3.451} \\ 
{0.49, 0.396, 1.144} &
 {0.49, 0.384, 6.635} &
 {0.31, 0.36, 3.025} &
 {0.351, 0.331, 5.002} &
 {0.349, 0.31, 2.025} &
 {0.373, 0.49, 4.072} &
 {0.396, 0.49, 4.598} &
 {0.39, 0.452, 5.039} &
 {0.49, 0.49, 5.347} &
 {0.31, 0.339, 3.447} &
 {0.49, 0.372, 6.646} &
 {0.49, 0.371, 3.149} &
 {0.31, 0.368, 1.613} &
 {0.49, 0.428, 3.313} &
 {0.323, 0.398, 3.204} \\ 
{0.38, 0.481, 0.974} &
 {0.31, 0.359, 5.886} &
 {0.49, 0.369, 6.069} &
 {0.489, 0.388, 4.257} &
 {0.42, 0.365, 4.347} &
 {0.359, 0.404, 1.914} &
 {0.369, 0.351, 1.909} &
 {0.395, 0.473, 4.492} &
 {0.404, 0.49, 4.157} &
 {0.392, 0.381, 2.102} &
 {0.393, 0.49, 5.302} &
 {0.421, 0.43, 8.049} &
 {0.31, 0.362, 2.433} &
 {0.49, 0.376, 5.196} &
 {0.49, 0.398, 2.915} \\ 
{0.49, 0.393, 3.448} &
 {0.49, 0.379, 5.126} &
 {0.31, 0.339, 3.783} &
 {0.388, 0.445, 4.7} &
 {0.372, 0.31, 2.221} &
 {0.376, 0.49, 1.441} &
 {0.389, 0.49, 4.666} &
 {0.384, 0.49, 5.486} &
 {0.31, 0.31, 7.37} &
 {0.49, 0.362, 3.614} &
 {0.49, 0.365, 1.555} &
 {0.31, 0.37, 0.109} &
 {0.49, 0.356, 4.067} &
 {0.31, 0.346, 2.265} &
 {0.328, 0.35, 7.346} \\ 
{0.49, 0.49, 3.521} &
 {0.398, 0.49, 3.22} &
 {0.378, 0.31, 3.174} &
 {0.365, 0.49, 4.994} &
 {0.358, 0.49, 2.614} &
 {0.328, 0.326, 4.208} &
 {0.462, 0.375, 2.709} &
 {0.49, 0.379, 2.394} &
 {0.49, 0.457, 7.808} &
 {0.49, 0.378, 5.963} &
 {0.318, 0.31, 5.289} &
 {0.381, 0.49, 1.089} &
 {0.401, 0.49, 4.785} &
 {0.419, 0.49, 5.794} &
 {0.451, 0.49, 1.945} \\ 

   \hline
    \end{tabular}}
        \caption{Solutions for 15 segments.}
    \label{table: sols for 15 segments.}
  \end{table}

  \begin{table}[htbp]
  \centering
  \resizebox{16cm}{!}{%
    \begin{tabular}{|c|c|c|c|c|c|c|c|c|c|c|c|c|c|c|c|}
   \hline
  {${w_a}_{0},{w_b}_{0},z_{0}$} [$\mu$m] & {${w_a}_{1},{w_b}_{1},z_{1}$} [$\mu$m] & {${w_a}_{2},{w_b}_{2},z_{2}$} [$\mu$m] & {${w_a}_{3},{w_b}_{3},z_{3}$} [$\mu$m] & {${w_a}_{4},{w_b}_{4},z_{4}$} [$\mu$m] & {${w_a}_{5},{w_b}_{5},z_{5}$} [$\mu$m] & {${w_a}_{6},{w_b}_{6},z_{6}$} [$\mu$m] & {${w_a}_{7},{w_b}_{7},z_{7}$} [$\mu$m] & {${w_a}_{8},{w_b}_{8},z_{8}$} [$\mu$m] & {${w_a}_{9},{w_b}_{9},z_{9}$} [$\mu$m] & {${w_a}_{10},{w_b}_{10},z_{10}$} [$\mu$m] & {${w_a}_{11},{w_b}_{11},z_{11}$} [$\mu$m] & {${w_a}_{12},{w_b}_{12},z_{12}$} [$\mu$m] & {${w_a}_{13},{w_b}_{13},z_{13}$} [$\mu$m] & {${w_a}_{14},{w_b}_{14},z_{14}$} [$\mu$m] & {${w_a}_{15},{w_b}_{15},z_{15}$} [$\mu$m]\\
   \hline
   {0.401, 0.31, 2.508} &
 {0.386, 0.49, 4.599} &
 {0.385, 0.49, 5.896} &
 {0.391, 0.394, 5.753} &
 {0.31, 0.333, 5.635} &
 {0.49, 0.364, 5.852} &
 {0.49, 0.378, 4.19} &
 {0.39, 0.37, 4.984} &
 {0.474, 0.487, 6.171} &
 {0.403, 0.454, 1.237} &
 {0.386, 0.49, 1.732} &
 {0.388, 0.49, 2.621} &
 {0.385, 0.31, 1.11} &
 {0.373, 0.49, 0.743} &
 {0.381, 0.31, 2.009} &
 {0.385, 0.49, 4.714} \\ 
{0.41, 0.49, 0.781} &
 {0.413, 0.49, 2.303} &
 {0.417, 0.49, 1.187} &
 {0.423, 0.49, 4.879} &
 {0.404, 0.452, 2.03} &
 {0.378, 0.395, 2.242} &
 {0.334, 0.338, 4.034} &
 {0.484, 0.391, 2.445} &
 {0.49, 0.397, 1.941} &
 {0.49, 0.395, 4.787} &
 {0.49, 0.386, 4.724} &
 {0.355, 0.364, 8.101} &
 {0.371, 0.472, 1.146} &
 {0.403, 0.49, 6.424} &
 {0.389, 0.49, 5.298} &
 {0.4, 0.351, 6.965} \\ 
{0.402, 0.49, 5.084} &
 {0.391, 0.432, 5.066} &
 {0.489, 0.48, 3.309} &
 {0.426, 0.406, 3.275} &
 {0.49, 0.441, 3.379} &
 {0.49, 0.374, 4.327} &
 {0.49, 0.413, 6.31} &
 {0.374, 0.345, 5.398} &
 {0.356, 0.432, 3.848} &
 {0.392, 0.444, 3.089} &
 {0.383, 0.469, 5.079} &
 {0.372, 0.469, 6.405} &
 {0.44, 0.392, 1.5} &
 {0.453, 0.403, 1.273} &
 {0.45, 0.387, 0.631} &
 {0.442, 0.354, 5.132} \\ 
{0.414, 0.49, 4.455} &
 {0.394, 0.49, 1.314} &
 {0.39, 0.49, 1.1} &
 {0.379, 0.49, 3.62} &
 {0.378, 0.31, 4.68} &
 {0.314, 0.342, 6.634} &
 {0.49, 0.349, 2.084} &
 {0.49, 0.359, 5.729} &
 {0.49, 0.353, 1.296} &
 {0.31, 0.333, 4.377} &
 {0.326, 0.31, 5.98} &
 {0.359, 0.49, 4.967} &
 {0.356, 0.49, 2.445} &
 {0.35, 0.49, 0.75} &
 {0.347, 0.49, 0.668} &
 {0.345, 0.31, 6.21} \\ 
{0.49, 0.437, 7.074} &
 {0.49, 0.416, 1.36} &
 {0.49, 0.413, 2.244} &
 {0.49, 0.41, 3.165} &
 {0.487, 0.397, 3.949} &
 {0.353, 0.426, 4.651} &
 {0.348, 0.489, 0.9} &
 {0.49, 0.49, 4.748} &
 {0.376, 0.49, 3.44} &
 {0.479, 0.49, 5.31} &
 {0.367, 0.49, 2.411} &
 {0.482, 0.49, 5.202} &
 {0.343, 0.382, 5.139} &
 {0.49, 0.414, 4.262} &
 {0.49, 0.389, 5.277} &
 {0.49, 0.434, 4.221} \\ 

   \hline
    \end{tabular}}
        \caption{Solutions for 16 segments.}
    \label{table: sols for 16 segments.}
  \end{table}

  \begin{table}[htbp]
  \centering
  \resizebox{16cm}{!}{%
    \begin{tabular}{|c|c|c|c|c|c|c|c|c|c|c|c|c|c|c|c|c|}
   \hline
  {${w_a}_{0},{w_b}_{0},z_{0}$} [$\mu$m] & {${w_a}_{1},{w_b}_{1},z_{1}$} [$\mu$m] & {${w_a}_{2},{w_b}_{2},z_{2}$} [$\mu$m] & {${w_a}_{3},{w_b}_{3},z_{3}$} [$\mu$m] & {${w_a}_{4},{w_b}_{4},z_{4}$} [$\mu$m] & {${w_a}_{5},{w_b}_{5},z_{5}$} [$\mu$m] & {${w_a}_{6},{w_b}_{6},z_{6}$} [$\mu$m] & {${w_a}_{7},{w_b}_{7},z_{7}$} [$\mu$m] & {${w_a}_{8},{w_b}_{8},z_{8}$} [$\mu$m] & {${w_a}_{9},{w_b}_{9},z_{9}$} [$\mu$m] & {${w_a}_{10},{w_b}_{10},z_{10}$} [$\mu$m] & {${w_a}_{11},{w_b}_{11},z_{11}$} [$\mu$m] & {${w_a}_{12},{w_b}_{12},z_{12}$} [$\mu$m] & {${w_a}_{13},{w_b}_{13},z_{13}$} [$\mu$m] & {${w_a}_{14},{w_b}_{14},z_{14}$} [$\mu$m] & {${w_a}_{15},{w_b}_{15},z_{15}$} [$\mu$m] & {${w_a}_{16},{w_b}_{16},z_{16}$} [$\mu$m]\\
   \hline
   {0.47, 0.354, 2.552} &
 {0.489, 0.489, 4.005} &
 {0.401, 0.489, 3.781} &
 {0.398, 0.49, 0.609} &
 {0.387, 0.49, 3.139} &
 {0.384, 0.49, 4.535} &
 {0.374, 0.332, 3.035} &
 {0.49, 0.49, 4.69} &
 {0.382, 0.377, 6.071} &
 {0.464, 0.369, 1.662} &
 {0.49, 0.376, 2.652} &
 {0.49, 0.373, 5.599} &
 {0.31, 0.31, 3.429} &
 {0.385, 0.403, 5.228} &
 {0.404, 0.49, 2.479} &
 {0.412, 0.49, 6.077} &
 {0.417, 0.49, 1.342} \\ 
{0.391, 0.49, 4.695} &
 {0.372, 0.49, 4.472} &
 {0.354, 0.319, 4.773} &
 {0.427, 0.49, 2.737} &
 {0.398, 0.381, 2.937} &
 {0.419, 0.38, 4.577} &
 {0.49, 0.385, 2.639} &
 {0.49, 0.387, 3.778} &
 {0.49, 0.388, 0.825} &
 {0.49, 0.405, 4.128} &
 {0.326, 0.334, 5.18} &
 {0.364, 0.49, 3.439} &
 {0.369, 0.49, 1.146} &
 {0.37, 0.31, 3.791} &
 {0.368, 0.49, 2.115} &
 {0.38, 0.49, 3.79} &
 {0.414, 0.418, 5.498} \\ 
{0.445, 0.31, 1.624} &
 {0.488, 0.49, 3.059} &
 {0.392, 0.49, 3.864} &
 {0.391, 0.49, 0.788} &
 {0.394, 0.49, 3.762} &
 {0.401, 0.49, 1.823} &
 {0.49, 0.49, 7.358} &
 {0.49, 0.49, 8.463} &
 {0.436, 0.379, 6.782} &
 {0.49, 0.379, 1.764} &
 {0.49, 0.379, 2.269} &
 {0.49, 0.374, 3.394} &
 {0.328, 0.351, 3.885} &
 {0.389, 0.313, 1.23} &
 {0.375, 0.49, 6.523} &
 {0.386, 0.31, 1.505} &
 {0.379, 0.49, 3.842} \\ 
{0.323, 0.417, 4.394} &
 {0.483, 0.34, 1.182} &
 {0.49, 0.358, 4.876} &
 {0.49, 0.49, 5.219} &
 {0.478, 0.354, 3.93} &
 {0.356, 0.403, 3.866} &
 {0.49, 0.49, 6.799} &
 {0.484, 0.49, 3.941} &
 {0.37, 0.49, 1.301} &
 {0.37, 0.49, 2.031} &
 {0.373, 0.49, 1.018} &
 {0.417, 0.49, 5.816} &
 {0.391, 0.445, 2.829} &
 {0.396, 0.39, 0.973} &
 {0.31, 0.333, 3.87} &
 {0.49, 0.379, 4.685} &
 {0.49, 0.387, 4.703} \\ 
{0.49, 0.364, 2.156} &
 {0.49, 0.355, 3.895} &
 {0.31, 0.325, 6.445} &
 {0.367, 0.49, 3.836} &
 {0.49, 0.49, 2.237} &
 {0.4, 0.49, 3.424} &
 {0.49, 0.49, 7.409} &
 {0.388, 0.49, 2.986} &
 {0.317, 0.391, 4.107} &
 {0.407, 0.351, 1.147} &
 {0.311, 0.354, 0.143} &
 {0.49, 0.368, 2.304} &
 {0.49, 0.379, 3.522} &
 {0.49, 0.389, 3.146} &
 {0.49, 0.403, 4.434} &
 {0.487, 0.49, 7.393} &
 {0.338, 0.49, 2.395} \\ 

   \hline
    \end{tabular}}
        \caption{Solutions for 17 segments.}
    \label{table: sols for 17 segments.}
  \end{table}

  \begin{table}[htbp]
  \centering
  \resizebox{16cm}{!}{%
    \begin{tabular}{|c|c|c|c|c|c|c|c|c|c|c|c|c|c|c|c|c|c|}
   \hline
  {${w_a}_{0},{w_b}_{0},z_{0}$} [$\mu$m] & {${w_a}_{1},{w_b}_{1},z_{1}$} [$\mu$m] & {${w_a}_{2},{w_b}_{2},z_{2}$} [$\mu$m] & {${w_a}_{3},{w_b}_{3},z_{3}$} [$\mu$m] & {${w_a}_{4},{w_b}_{4},z_{4}$} [$\mu$m] & {${w_a}_{5},{w_b}_{5},z_{5}$} [$\mu$m] & {${w_a}_{6},{w_b}_{6},z_{6}$} [$\mu$m] & {${w_a}_{7},{w_b}_{7},z_{7}$} [$\mu$m] & {${w_a}_{8},{w_b}_{8},z_{8}$} [$\mu$m] & {${w_a}_{9},{w_b}_{9},z_{9}$} [$\mu$m] & {${w_a}_{10},{w_b}_{10},z_{10}$} [$\mu$m] & {${w_a}_{11},{w_b}_{11},z_{11}$} [$\mu$m] & {${w_a}_{12},{w_b}_{12},z_{12}$} [$\mu$m] & {${w_a}_{13},{w_b}_{13},z_{13}$} [$\mu$m] & {${w_a}_{14},{w_b}_{14},z_{14}$} [$\mu$m] & {${w_a}_{15},{w_b}_{15},z_{15}$} [$\mu$m] & {${w_a}_{16},{w_b}_{16},z_{16}$} [$\mu$m] & {${w_a}_{17},{w_b}_{17},z_{17}$} [$\mu$m]\\
   \hline
   {0.386, 0.49, 0.834} &
 {0.403, 0.483, 6.047} &
 {0.457, 0.397, 3.952} &
 {0.49, 0.392, 4.356} &
 {0.452, 0.381, 4.493} &
 {0.49, 0.38, 1.682} &
 {0.49, 0.37, 2.574} &
 {0.31, 0.353, 3.029} &
 {0.322, 0.315, 5.32} &
 {0.356, 0.49, 4.989} &
 {0.361, 0.49, 0.483} &
 {0.358, 0.49, 3.438} &
 {0.343, 0.334, 2.031} &
 {0.398, 0.31, 1.912} &
 {0.326, 0.454, 1.552} &
 {0.393, 0.436, 4.469} &
 {0.478, 0.38, 4.728} &
 {0.444, 0.38, 5.29} \\ 
{0.35, 0.451, 5.791} &
 {0.432, 0.38, 4.761} &
 {0.49, 0.401, 4.322} &
 {0.49, 0.401, 3.771} &
 {0.49, 0.405, 1.201} &
 {0.488, 0.405, 1.511} &
 {0.49, 0.49, 8.36} &
 {0.367, 0.335, 5.084} &
 {0.372, 0.49, 1.354} &
 {0.378, 0.49, 1.564} &
 {0.381, 0.49, 1.55} &
 {0.384, 0.49, 3.783} &
 {0.384, 0.49, 1.822} &
 {0.365, 0.426, 4.81} &
 {0.31, 0.31, 1.859} &
 {0.448, 0.375, 5.089} &
 {0.49, 0.391, 4.158} &
 {0.486, 0.388, 0.078} \\ 
{0.398, 0.49, 4.423} &
 {0.387, 0.49, 4.296} &
 {0.359, 0.34, 2.057} &
 {0.335, 0.31, 2.351} &
 {0.409, 0.328, 1.224} &
 {0.446, 0.343, 0.586} &
 {0.49, 0.487, 6.681} &
 {0.49, 0.438, 3.391} &
 {0.49, 0.388, 4.126} &
 {0.49, 0.379, 4.462} &
 {0.31, 0.338, 5.335} &
 {0.358, 0.31, 2.155} &
 {0.368, 0.49, 3.334} &
 {0.371, 0.49, 1.74} &
 {0.366, 0.49, 1.802} &
 {0.358, 0.49, 3.321} &
 {0.355, 0.31, 4.672} &
 {0.406, 0.354, 1.982} \\ 
{0.406, 0.49, 5.178} &
 {0.424, 0.49, 5.531} &
 {0.437, 0.444, 5.113} &
 {0.433, 0.393, 2.418} &
 {0.49, 0.386, 3.994} &
 {0.49, 0.37, 3.713} &
 {0.49, 0.358, 3.176} &
 {0.31, 0.381, 3.249} &
 {0.456, 0.345, 0.478} &
 {0.448, 0.385, 2.873} &
 {0.489, 0.49, 2.385} &
 {0.388, 0.311, 0.058} &
 {0.363, 0.398, 4.681} &
 {0.365, 0.49, 1.23} &
 {0.364, 0.49, 5.676} &
 {0.373, 0.31, 2.632} &
 {0.357, 0.434, 5.507} &
 {0.49, 0.348, 3.767} \\ 
{0.381, 0.31, 4.163} &
 {0.362, 0.49, 3.043} &
 {0.37, 0.49, 3.352} &
 {0.368, 0.31, 3.492} &
 {0.364, 0.49, 4.538} &
 {0.37, 0.49, 0.91} &
 {0.334, 0.358, 5.803} &
 {0.484, 0.365, 0.655} &
 {0.49, 0.384, 1.45} &
 {0.49, 0.386, 1.45} &
 {0.49, 0.392, 2.721} &
 {0.49, 0.391, 5.346} &
 {0.49, 0.378, 2.243} &
 {0.31, 0.316, 6.75} &
 {0.348, 0.49, 0.899} &
 {0.356, 0.31, 2.903} &
 {0.377, 0.49, 4.269} &
 {0.391, 0.49, 3.061} \\ 

   \hline
    \end{tabular}}
        \caption{Solutions for 18 segments.}
    \label{table: sols for 18 segments.}
  \end{table}

  \begin{table}[htbp]
  \centering
  \resizebox{16cm}{!}{%
    \begin{tabular}{|c|c|c|c|c|c|c|c|c|c|c|c|c|c|c|c|c|c|c|}
   \hline
  {${w_a}_{0},{w_b}_{0},z_{0}$} [$\mu$m] & {${w_a}_{1},{w_b}_{1},z_{1}$} [$\mu$m] & {${w_a}_{2},{w_b}_{2},z_{2}$} [$\mu$m] & {${w_a}_{3},{w_b}_{3},z_{3}$} [$\mu$m] & {${w_a}_{4},{w_b}_{4},z_{4}$} [$\mu$m] & {${w_a}_{5},{w_b}_{5},z_{5}$} [$\mu$m] & {${w_a}_{6},{w_b}_{6},z_{6}$} [$\mu$m] & {${w_a}_{7},{w_b}_{7},z_{7}$} [$\mu$m] & {${w_a}_{8},{w_b}_{8},z_{8}$} [$\mu$m] & {${w_a}_{9},{w_b}_{9},z_{9}$} [$\mu$m] & {${w_a}_{10},{w_b}_{10},z_{10}$} [$\mu$m] & {${w_a}_{11},{w_b}_{11},z_{11}$} [$\mu$m] & {${w_a}_{12},{w_b}_{12},z_{12}$} [$\mu$m] & {${w_a}_{13},{w_b}_{13},z_{13}$} [$\mu$m] & {${w_a}_{14},{w_b}_{14},z_{14}$} [$\mu$m] & {${w_a}_{15},{w_b}_{15},z_{15}$} [$\mu$m] & {${w_a}_{16},{w_b}_{16},z_{16}$} [$\mu$m] & {${w_a}_{17},{w_b}_{17},z_{17}$} [$\mu$m] & {${w_a}_{18},{w_b}_{18},z_{18}$} [$\mu$m]\\
   \hline
   {0.49, 0.4, 4.349} &
 {0.49, 0.389, 2.291} &
 {0.417, 0.385, 4.477} &
 {0.31, 0.406, 1.322} &
 {0.49, 0.49, 5.257} &
 {0.434, 0.439, 0.072} &
 {0.489, 0.49, 2.748} &
 {0.35, 0.312, 3.41} &
 {0.35, 0.49, 3.24} &
 {0.36, 0.49, 3.152} &
 {0.362, 0.49, 2.108} &
 {0.353, 0.49, 1.3} &
 {0.31, 0.31, 6.955} &
 {0.49, 0.36, 4.382} &
 {0.49, 0.36, 2.885} &
 {0.49, 0.361, 0.109} &
 {0.49, 0.354, 1.63} &
 {0.31, 0.358, 3.862} &
 {0.49, 0.479, 5.951} \\ 
{0.399, 0.49, 3.586} &
 {0.379, 0.31, 3.419} &
 {0.372, 0.424, 3.959} &
 {0.38, 0.489, 0.106} &
 {0.384, 0.49, 1.941} &
 {0.389, 0.49, 1.177} &
 {0.397, 0.49, 3.265} &
 {0.37, 0.403, 5.534} &
 {0.447, 0.38, 4.612} &
 {0.49, 0.394, 4.986} &
 {0.49, 0.38, 5.121} &
 {0.49, 0.363, 1.052} &
 {0.321, 0.334, 6.85} &
 {0.372, 0.311, 0.116} &
 {0.365, 0.31, 1.484} &
 {0.381, 0.49, 1.419} &
 {0.395, 0.49, 3.738} &
 {0.407, 0.49, 3.415} &
 {0.45, 0.49, 4.444} \\ 
{0.49, 0.49, 4.991} &
 {0.31, 0.372, 2.712} &
 {0.31, 0.36, 0.872} &
 {0.49, 0.354, 1.585} &
 {0.49, 0.362, 3.494} &
 {0.49, 0.363, 4.809} &
 {0.338, 0.407, 4.586} &
 {0.354, 0.31, 4.725} &
 {0.367, 0.49, 4.491} &
 {0.373, 0.49, 4.882} &
 {0.359, 0.385, 3.538} &
 {0.359, 0.379, 2.709} &
 {0.317, 0.36, 0.986} &
 {0.414, 0.401, 2.937} &
 {0.485, 0.42, 3.03} &
 {0.49, 0.417, 2.442} &
 {0.49, 0.413, 3.516} &
 {0.49, 0.416, 1.446} &
 {0.49, 0.423, 1.769} \\ 
{0.409, 0.49, 1.799} &
 {0.388, 0.432, 1.162} &
 {0.385, 0.49, 3.01} &
 {0.427, 0.476, 3.994} &
 {0.355, 0.423, 3.937} &
 {0.49, 0.406, 4.782} &
 {0.49, 0.381, 4.5} &
 {0.49, 0.391, 5.05} &
 {0.475, 0.404, 0.292} &
 {0.466, 0.365, 4.909} &
 {0.338, 0.441, 4.836} &
 {0.421, 0.49, 2.683} &
 {0.449, 0.49, 1.016} &
 {0.414, 0.49, 5.068} &
 {0.362, 0.476, 3.954} &
 {0.434, 0.31, 2.387} &
 {0.49, 0.49, 2.232} &
 {0.49, 0.435, 3.007} &
 {0.471, 0.49, 4.867} \\ 

   \hline
    \end{tabular}}
        \caption{Solutions for 19 segments.}
    \label{table: sols for 19 segments.}
  \end{table}

\end{document}